\documentclass[aps,pre,twocolumn,superscriptaddress,amsmath,amssymb,noshowpacs,longbibliography]{revtex4-2}   
\usepackage[pdftex]{graphics}
\usepackage{epsfig,epstopdf}
\usepackage{dcolumn}
\usepackage{bm}
\usepackage{color}
\usepackage{hyperref} 

\usepackage{tikz}   
\usetikzlibrary{arrows}  
\usepackage{pgfplots}
\pgfplotsset{compat=1.14} 

\pgfrealjobname{ADR_article} 

\def\clap#1{\hbox to 0pt{\hss#1\hss}}

\newcommand{\bx}{{\mathbf{x}}}
\newcommand{\bk}{{\mathbf{k}}}
\newcommand{\be}{{\mathbf{e}}}
\newcommand{\bl}{{\mathbf{l}}}
\newcommand{\bq}{{\mathbf{q}}}                     \newcommand{\ba}{{\mathbf{a}}}

\begin{document}

\title{Rectangle--triangle soft-matter quasicrystals with hexagonal symmetry}
\author{Andrew J. Archer}
\email{A.J.Archer@lboro.ac.uk}
\affiliation{Department of Mathematical Sciences and Interdisciplinary Centre for Mathematical Modelling, Loughborough University, Loughborough, Leicestershire LE11 3TU, United Kingdom}
\author{Tomonari Dotera}
\email{dotera@phys.kindai.ac.jp}
\affiliation{Department of Physics, Kindai University, 3-4-1 Kowakae Higashi-Osaka 577-8502, Japan}
\author{Alastair M. Rucklidge}
\email{A.M.Rucklidge@leeds.ac.uk}
\affiliation{School of Mathematics, University of Leeds, Leeds LS2 9JT, United Kingdom}

\begin{abstract}
Aperiodic (quasicrystalline) tilings, such as Penrose's tiling, can be built up from e.g.\ kites and darts, squares
and equilateral triangles, rhombi or shield shaped tiles and can have a variety of different symmetries. However,
almost all quasicrystals occurring in soft-matter are of the dodecagonal type. Here, we investigate a class of
aperiodic tilings with hexagonal symmetry that are based on rectangles and two types of equilateral triangles. We show
how to design soft-matter systems of particles interacting via pair potentials containing two length-scales that form
aperiodic stable states with two different examples of rectangle--triangle tilings. One of these is the bronze-mean
tiling, while the other is a generalization. Our work points to how more general (beyond dodecagonal) quasicrystals
can be designed in soft-matter.
 \end{abstract}
\date{\today}

\maketitle 

\section{Introduction}

In the `game' of arranging tiles in a plane, one of the more fascinating and striking things to emerge are quasicrystals (QCs), which lack the usual spatial periodicity of `simple' tilings. The classic example is the Penrose tiling, formed e.g.\ of rhombi with 36$^\circ$ and 72$^\circ$ corner angles \cite{Baake2013}. Such patterns have long been of interest due to their aesthetic and mathematical beauty. Shechtman's 1982 discovery \cite{Shechtman1984a}, confirmed and built upon now in a sizable body of work, shows that in nature, atoms, molecules, nanoparticles and polymeric soft-matter are capable of self-assembling into such structures~\cite{Zeng2004, Hayashida2007, Talapin2009, Dotera2011, Fischer2011, Xiao2012, Ishimasa2011, Forster2016, Ye2017, Jayaraman2021}. A characteristic feature of QCs, and of aperiodic tilings, is that they have sharp Bragg peaks in their diffraction patterns, or equivalently, their Fourier transforms are point spectra. They often also have unusual (e.g.\ icosahedral, ten- or twelve-fold) rotation symmetries that preclude spatial periodicity.

\begin{figure}[t]
\begin{center}
\includegraphics{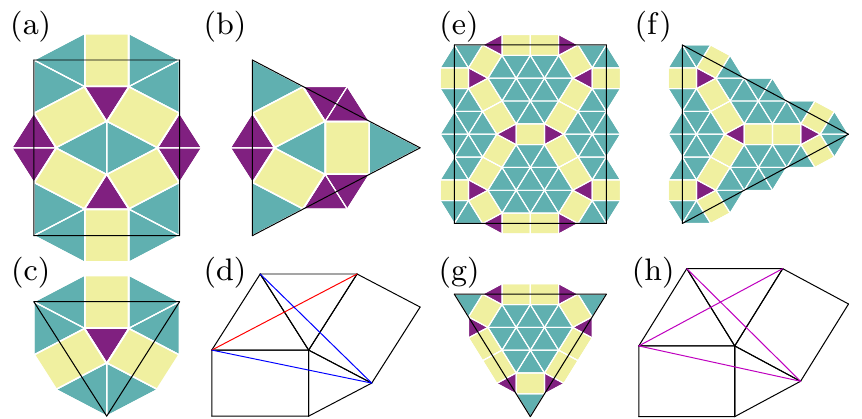}
\end{center}
\caption{\label{fig:BM_EDRT_inflation}
Subdivision schemes of the BM (a)--(c) and EDRT (e)--(g) tilings. 
Rectangles are yellow, small triangles are purple and large triangles are green.
The colored tiles that protrude from the inflated tiles overlap when the larger tiles are joined. 
In the BM tiling~(d), the diagonal across two large triangles (red line) is longer than the diagonals across the small (or large) triangles and the rectangle (blue lines). 
In the EDRT tiling~(f), these diagonals (magenta lines) are equal.}
\end{figure}

Recently, Dotera et al.~\cite{Dotera2017} discovered another striking aperiodic tiling, but with six-fold rotation symmetry. 
It is known as the bronze mean (BM) tiling, and is formed from rectangles and two different sizes of equilateral triangles (see Fig.~\ref{fig:BM_EDRT_inflation}). 
They also showed how particles having a hard core and a repulsive shoulder can self-assemble into this structure. 
Soft matter systems have also been observed to form structures that can be described by a rectangle--triangle tilings~\cite{Ishimasa2011, Forster2016, Ye2017, Jayaraman2021}. 
There is rich geometry and beauty in the BM tiling. It is the third member of a family of `metallic mean' tilings that are associated with the irrational roots of the quadratic $x^2 - mx - 1 = 0$, where $m$~is a positive integer. 
The Penrose tiling~\cite{Penrose1974, Baake2013}, with five-fold symmetry, features the golden mean (GM), $\frac{1}{2}(1+\sqrt{5})\approx1.618$ as a characteristic ratio in the $m=1$ case. The Ammann--Beenker tiling~\cite{Baake2013}, with eight-fold symmetry, has the silver mean (SM), $1+\sqrt{2}\approx2.414$, for $m=2$. The $m=3$ bronze mean is $\frac{1}{2}(3+\sqrt{13})\approx3.303$, and the BM tiling has 6-fold rotational symmetry, but like all QCs, no spatial periodicity. 
The three aperiodic tilings just mentioned all have Fourier spectra exhibiting primary peaks distributed around a circle, with ten peaks in the GM (Penrose) case, eight in the SM case and twelve in the BM case. 
However, the full spectra are dense, i.e.\ there are Bragg peaks arbitrarily close to any point in Fourier $\bk$-space.

Associating the average position of thermal particles with locations in each of these tiles provides a natural way to describe the structure of quasicrystalline materials. The (probability) density distribution $\rho(\bx)$ of these particles is a continuous field and the Fourier transform of $\rho(\bx)$ in a QC exhibits the same features: a dense set of Bragg peaks and the same rotation symmetries. 

This provides a link to applying the methods of pattern formation theory (PFT) to these tiling-generated structures.
Here, starting from the BM tiling spectrum and the idea that having two lengthscales can stabilize
quasicrystals~\cite{Lifshitz2007a}, we identify relevant circles in the Fourier spectrum and develop a soft-core
particle model that has a stable QC density profile with the BM structure. Building this bridge between tilings and
PFT gives insight into other related tilings, one of which, related to the family of tilings introduced in Nakakura et
al.~\cite{Nakakura2019}, we present and investigate here.

The central objects of study in PFT are usually partial differential equations (PDEs), and there are many powerful ideas, such as nonlinear mode interactions, for understanding the emergence and stability of patterns in a wide range of problems~\cite{Cross1993, Hoyle2006}. The (integro-)PDEs we consider here come from statistical mechanics, in particular dynamical density functional theory (DDFT)~\cite{Marconi1999, Archer2004a, Hansen2013, teVrugt2020}, which is a theory for the time-evolution of~$\rho(\bx)$.

This paper is arranged as follows: in Sec.~\ref{sec:2} we describe the bronze mean and equidiagonal rectangle--triangle tilings, illustrate the inflation rules for constructing them and the corresponding Fourier power spectra that we then use to bridge between these tilings and \hbox{PFT}. 
Then, in Sec.~\ref{sec:3} we briefly explain how PFT allows us to identify two circles on the Fourier spectra and to determine which Fourier modes are needed to form the tiling structures. 
In Sec.~\ref{sec:4} we describe how to go from the Fourier spectrum for each tiling to determine in each case a soft particle model with pair interactions that make the corresponding tilings stable. 
We use DDFT to obtain density profiles~$\rho(\bx)$, and to demonstrate that the resulting structures are stable.
However, these profiles are local minima of the free energy and not the global minima of the free energy, 
so they are not thermodynamically stable, only metastable states.
We also show how to match a subset of the maxima in~$\rho(\bx)$ with the vertices of the corresponding tiling. 
In Sec.~\ref{sec:5} we give further details of how to calculate the density profiles, in particular showing what size of box in which to calculate~$\rho(\bx)$. 
Since QCs have no unit cell, often one must resort to calculating~$\rho(\bx)$ on a finite-size domain with periodic boundary conditions, thus actually obtaining an {\em approximant} to the true~\hbox{QC}. 
We show how to select the box size so as to minimize errors from working with a finite size piece of the~\hbox{QC}. 
We also discuss how this approach is related to some other possible approaches to constructing periodic approximants. 
In Sec.~\ref{sec:6} we describe some of the key characteristic properties of the new equidiagonal rectangle--triangle tiling, including the inflation factor, numbers of the different tiles and the projection window. 
Finally, in Sec.~\ref{sec:7} we make a few concluding remarks.

\begin{figure}[t]
\begin{center}
\includegraphics{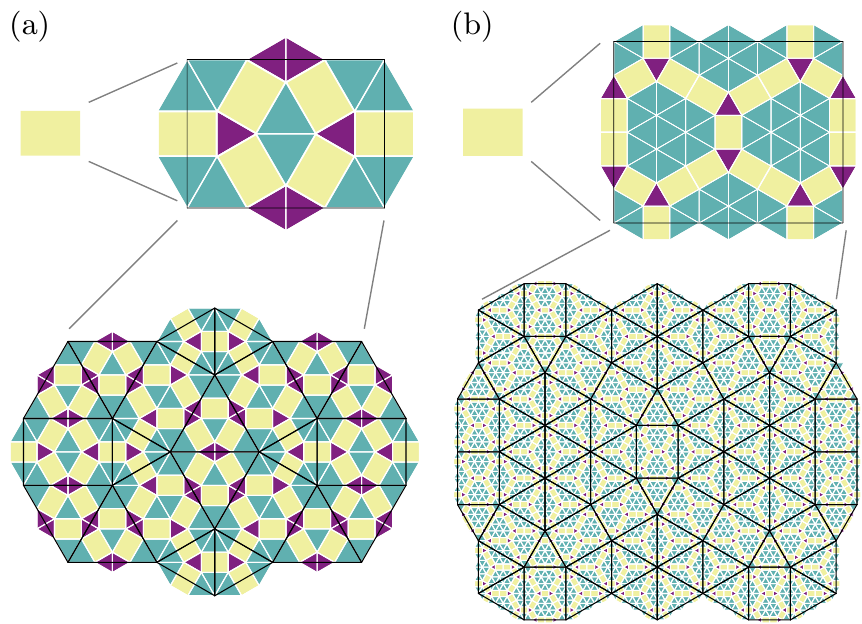}
\end{center}
\caption{\label{fig:BM_EDRT_R_example}
A single rectangle is inflated twice using the rules in Fig.~\ref{fig:BM_EDRT_inflation}; (a)~is the BM case and (b)~the~\hbox{EDRT}.}
\end{figure}

\section{Bronze Mean and Equidiagonal Rectangle--Triangle tilings}
\label{sec:2}

In Fig.~\ref{fig:BM_EDRT_inflation}(a)--(c), we illustrate the set of three tiles from which the BM tiling is built up: a rectangle and two equilateral triangles whose sides are the lengths of the two sides of the rectangle. The ratio of the sides of the rectangle is $\tfrac{1}{6}(\sqrt{3}+\sqrt{39})\approx1.330$. The aperiodic tiling is created using an inflation rule [illustrated in Fig.~\ref{fig:BM_EDRT_R_example}(a)], as described in~\cite{Dotera2017,Nakakura2019}. The inflation scaling factor is the Bronze Mean, $\tfrac{1}{2}(3+\sqrt{13})\approx3.303$.

In the BM tiling, there are two lengths that are almost the same: the diagonal across two large triangles, and the diagonal across the small (or large) triangle and the rectangle (see Fig.~\ref{fig:BM_EDRT_inflation}(d), red and blue lines). We change the ratio of the sides of the rectangle slightly, to $(\sqrt{3}+\sqrt{11})/4\approx1.262$, to make these two diagonals equal in length (see Fig.~\ref{fig:BM_EDRT_inflation}(h), magenta lines), which leads to a new aperiodic tiling with tiling subdivision rule shown in Fig.~\ref{fig:BM_EDRT_inflation}(e)--(g) and the inflation illustrated in Fig.~\ref{fig:BM_EDRT_R_example}(b), with a much larger inflation factor of $2\sqrt{3}+\sqrt{11}\approx6.781$. We refer to this as the equidiagonal rectangle--triangle (EDRT) tiling. As can be seen from Fig.~\ref{fig:BM_EDRT_R_example}(b) [see also Fig.~\ref{fig:BM_EDRT_inflation}(e)--(g)], the larger inflation factor corresponds to a much larger number of tiles in each inflated tile. Note also the patches of large triangles are bigger in this striking structure and also that the rectangles rotate by 90$^\circ$ after each inflation, leading to a non-Pisot inflation factor.

\begin{figure}
\begin{center}
\includegraphics{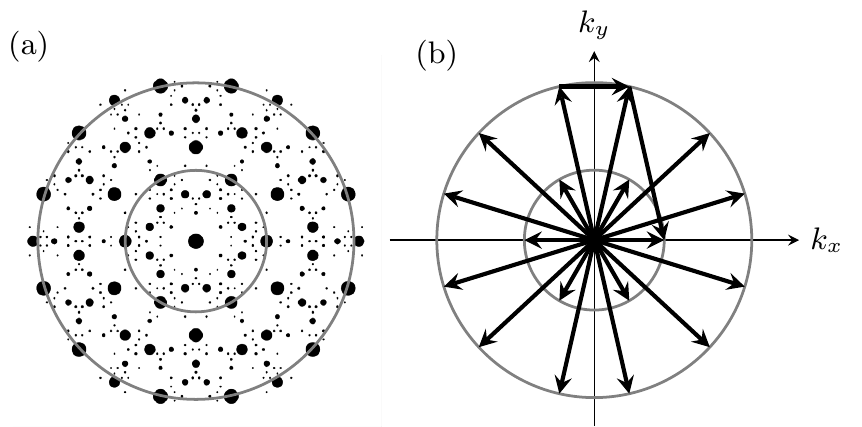}
\includegraphics{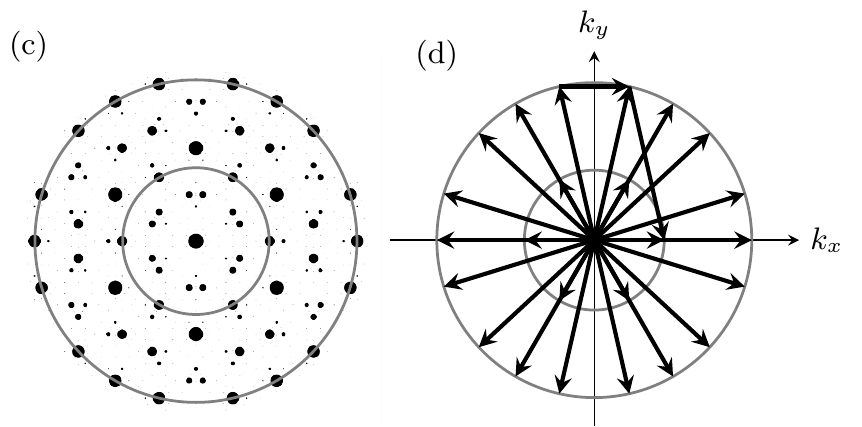}
\end{center}
\caption{(a,c) Power spectrum of aperiodic tiling and (b,d)~wave vectors, with (a,b)~BM and (c,d)~\hbox{EDRT}. The short vectors are of length $k_1=1$ while the longer have length $k_2\approx2.252$ (BM) and $k_2\approx2.186$ (EDRT). Pairs of the long vectors add to give the small ones. Making such Fourier triangles energetically favourable in the particle systems aids in stabilizing these structures.}
\label{fig:Spectra_and_wavevectors}
\end{figure}

In Fig.~\ref{fig:Spectra_and_wavevectors}(a) and (c) we show the Fourier power spectra of the BM and the EDRT tilings, respectively. 
These are obtained by forming the tiling as projection from a four-dimensional periodic structure and then calculating the Fourier transform of a large but finite portion of this projection, displaying only peaks with intensity greater than 0.0045 times that of the central peak. 
Further information about these tilings and their Fourier spectra are given in Sec.~\ref{sec:6}; see also Ref.~\cite{Dotera2017}.

\section{Pattern formation theory}
\label{sec:3}

We next call on ideas from PFT to select two circles in Fourier space: these correspond to two length-scales that we build into the pair interaction potentials between soft particles. 
Stabilizing patterns in continuum models involves nonlinear interactions between density waves acting to reinforce each other~\cite{Alexander1978, Bak1985, Lifshitz1997, Rucklidge2012, Archer2013, Ratliff2019, Castelino2020}. 
Two modes interact with a third when the wavevectors add up, as illustrated in Fig.~\ref{fig:Spectra_and_wavevectors}(b) and (d), and typically, in problems of minimising a free energy, having more three-wave interactions leads to an enhancement of the stability of the structure containing all these waves~\cite{Lifshitz1997}. 

The choice of circles is not unique, and we used three criteria to select the circles.
First, the peaks lying on the two circles should be amongst the strongest in the power spectrum of the tiling, so the resulting pattern should stand a good chance of resembling the tiling.
Second, there should be three-wave interactions: two vectors on one of the circles should add up to a vector on the other, required for stabilising the quasiperiodic pattern.
And third, the ratio between the two radii should be greater than two, which implies that the three-wave interaction must be two long vectors on the outer circle adding up to a small vector on the inner.
In the simplest cases of hexagonal symmetry, this criterion also implies that there should be twelve vectors (with uneven spacing) on the outer circle and six on the inner.
The reason for this third criterion is that it simplifies the possible three-wave interactions: with a radius ratio less than two, two vectors on the inner circle could add up to a vector on the outer, which would result in the complication of competition between two types of pattern~\cite{Rucklidge2012, Castelino2020}. 

The pairs of circles we select are displayed in Fig.~\ref{fig:Spectra_and_wavevectors}(a) and (c). 
In the EDRT tiling, the six peaks just off the outer circle in the BM spectrum have moved onto the outer circle, making a total of eighteen peaks.

In the following section we discuss systems of interacting soft particles, treated using DDFT, thus allowing us to
incorporate these ideas from PFT to tune the interactions between particles so that they are stable in either the BM
or EDRT structures.

Note also that to calculate the density field $\rho(\bx)$ from DDFT for a QC a periodic approximation is necessary. 
The considerations required to do this are discussed in Sec.~\ref{sec:5}.

\section{Dynamical density functional theory}
\label{sec:4}

Having found the specific favourable wavenumbers and Fourier modes for forming these structures, we can then identify the soft-matter systems in which the interactions between the particles leads to these modes being prominent in $\rho(\bx)$, i.e.\ where the free energy is lowered by modes on these two circles having a large amplitude. 
As mentioned, DDFT is a theory for the time evolution of $\rho(\bx)$, with the dynamics given by \cite{Marconi1999,Archer2004a,Hansen2013,teVrugt2020}
\begin{equation}
 \frac{\partial \rho}{\partial t}=\nabla\cdot\left[ \Gamma \rho\nabla\frac{\delta F[\rho]}{\delta\rho}\right]
 \label{eq:DDFT}
\end{equation}
where $\Gamma=D/k_BT$, $D$ is the diffusion coefficient, $T$ is the temperature and $k_B$ is Boltzmann's constant. Also, $F[\rho]$ is the Helmholtz free energy functional from {\em equilibrium} density functional theory (DFT) \cite{Hansen2013,Evans1979a},
\begin{equation}
F[\rho]=k_BT\int d \bx\, \rho[\ln \Lambda^d\rho-1]+\int d \bx\, U\rho+F_{ex}[\rho].
 \label{eq:F}
\end{equation}
The first term is the ideal-gas contribution ($\Lambda$ is the thermal de Broglie wavelength and $d$ is the dimensionality of the system), the second is the contribution from any external potential $U(\bx)$ (here $U=0$) and the last term is the contribution from the interactions between the particles. For soft particles interacting via a pair potential $v(r)$ that is finite for all inter-particles distances $r$, the following simple approximation is rather accurate \cite{Likos2001, Archer2014, Walters2018}
\begin{equation}
F_{ex}[\rho]=\frac{1}{2} \int d\bx \int d\bx' \rho(\bx)\rho(\bx')v(|\bx-\bx'|)
 \label{eq:F_ex}
\end{equation}
and so is used here. This can be re-written as the Fourier space integral
\begin{equation}
F_{ex}[\rho]=\frac{1}{2(2\pi)^d} \int d\bk |\hat{\rho}(\bk)|^2\hat{v}(k),
 \label{eq:F_ex_Fourier}
\end{equation}
where $\hat{\rho}(\bk)=\int d\bx \rho(\bx)e^{-i\bk\cdot\bx}$ is the Fourier transform of the density profile and $\hat{v}(k)$ is the Fourier transform of the pair potential, with $k=|\bk|$. From Eq.~\eqref{eq:F_ex_Fourier}, one can see that modes $\hat{\rho}(\bk)$ that correspond to {\em minima} in $\hat{v}(k)$ decrease $F_{ex}$ and so are likely to be favourable. By performing a linear stability analysis of the full equation~\eqref{eq:DDFT} for the uniform liquid of density~$\bar\rho$, one can determine the dispersion relation, which gives the growth or decay rate of modes with wavenumber $k$ and also provides a measure of the relative contributions of the energetic $F_{ex}[\rho]$ and the entropic (ideal-gas) parts to the free energy. The resulting criterion for marginal stability at wavenumber~$k_c$ is $1+{\bar\rho}\beta{\hat v}(k_c)=0$, where $\beta=(k_BT)^{-1}$ \cite{Archer2012, Archer2013, Ratliff2019}.

Thus, our approach here is to construct pair potentials~$v(r)$ so that the liquid is marginally stable at the two wavenumbers $k_1$ and~$k_2$ (radii of the two Fourier space circles in Fig.~\ref{fig:Spectra_and_wavevectors}) identified from our analysis of the tilings, as described in Sec.~\ref{sec:3}.
We scale lengths so that the smaller wavenumber is $k_1=1$, and then for the BM tiling the larger wavenumber is $k_2\approx2.252$ and for the EDRT tiling, $k_2\approx2.186$. 
We achieve this by using the following pair potential
\begin{equation}
v(r)=\varepsilon e^{-\tfrac{1}{2}\sigma^2 r^2}
\left(1+C_2r^2+C_4r^4+C_6r^6+C_8r^8\right)
\label{eq:v}
\end{equation}
that was originally proposed by Barkan et al.~\cite{Barkan2014}. Note that other soft-core systems also form QCs~\cite{Barkan2011,Archer2013,Zu2017,Scacchi2020,Malescio2021}. 
Soft-core potentials arise as the coarse-grained effective potentials between polymeric macromolecules in solution, such as star polymers, dendrimers or block copolymers \cite{Likos2001}. 
The parameter $\varepsilon$ in Eq.~\eqref{eq:v} controls the overall strength of $v(r)$, while the others ${\cal P}=\{\sigma,C_2,C_4,C_6,C_8\}$ can be chosen to determine the location and sharpness of the two minima in~$\hat{v}(k)$.
Choosing $\bar\rho=10$ and $\beta\varepsilon=1$, and requiring that the liquid be marginally stable at $k=k_1$ and $k=k_2$ results in four relations between the five parameters~$\cal P$. 
In practice, we choose the value of~$\sigma$ and find the other four parameters using these relations. 
A similar approach is discussed in Ref.~\cite{Ratliff2019} (see also \cite{Barkan2014, Barkan2015}). 

\begin{figure}
\begin{center}
\includegraphics{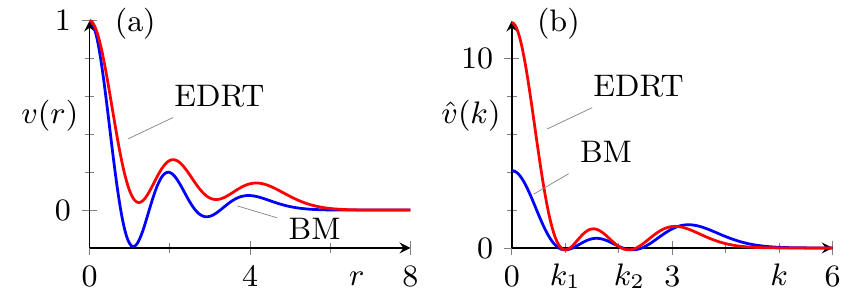}
\end{center}
    \caption{(a) Interaction potential~$v(r)$ and (b)~its Fourier transform~$\hat{v}(k)$ for the BM tiling (blue) and EDRT tiling (red) cases. The minima in the ${\hat v}(k)$ curves occur at $k_1=1$ (both cases) and at $k_2\approx2.252$ (BM) and $k_2\approx2.186$ (EDRT).}
    \label{fig:V_and_Vhat}
\end{figure}

The choice of~$\sigma$ is not entirely straight-forward since for the desired structures to be stable, one must at the same time make sure that the value of $\hat{v}(k)$ is sufficiently high at other wavenumbers~$k$ corresponding to competing crystal structures~\cite{Ratliff2019}. 
We determined~$\sigma$ largely by trial and error, using insight from the rate of convergence (or divergence) of the Picard-iteration algorithm used to solve the DDFT~\cite{Roth2010}. 
We find ${\cal P}={\cal P}_{BM}\equiv\{0.95, -2.3455, 1.2638, -0.20667, 0.010136\}$ leads to a stable BM structure. Similarly, ${\cal P}={\cal P}_{EDRT}\equiv\{0.85, -1.4516, 0.69075, -0.098040, 0.0044548\}$ leads to a stable EDRT structure. The potentials~$v(r)$ and their Fourier transforms ${\hat v}(k)$ are displayed in Fig.~\ref{fig:V_and_Vhat}.

\begin{figure*}
    \centering
    \includegraphics[width=0.90\hsize]{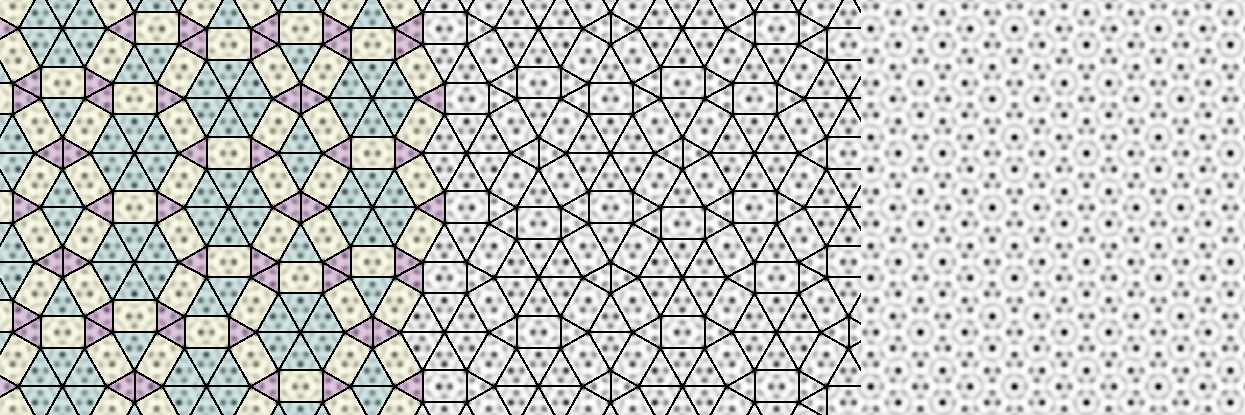}

    \includegraphics[width=0.90\hsize]{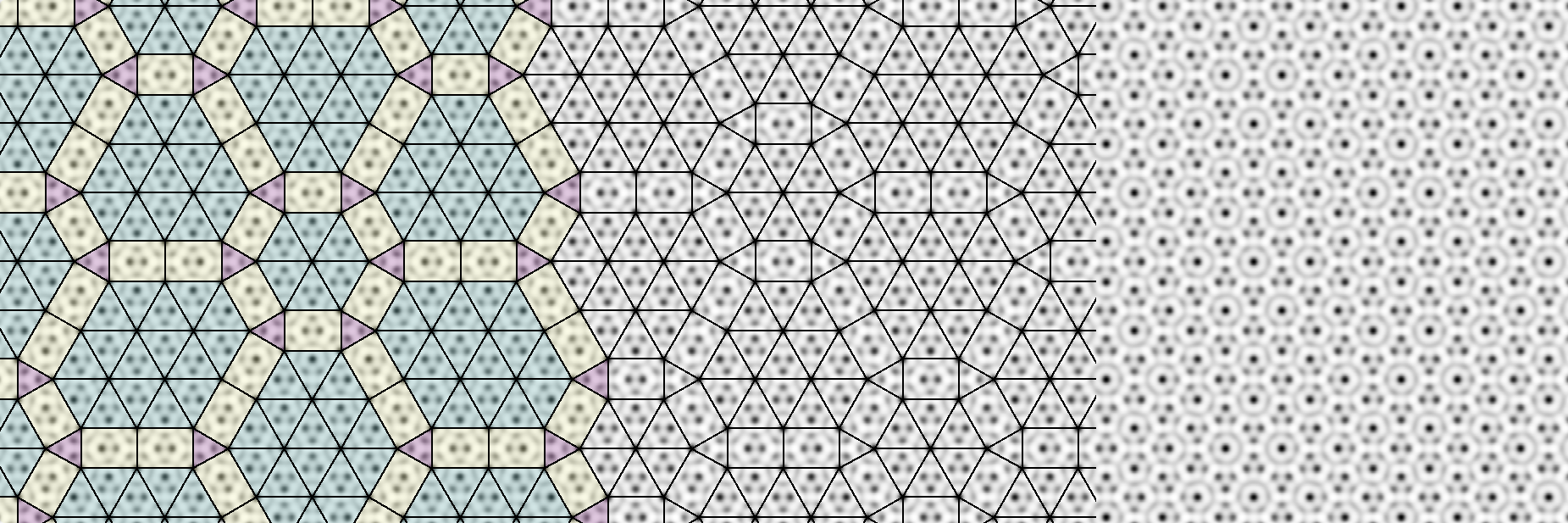}
    \caption{Grayscale plots of density profiles for the BM (top) and EDRT (bottom) systems, where black corresponds
    to maxima, where $\rho\approx25$, while white corresponds to minima, where $\rho\approx2$. Superimposed on each is
    the corresponding tiling. In the BM case, the short and long edges are of length 8.36 and 11.61, while in the EDRT
    case, these lengths are 8.72 and 11.95. We display here only small portions of profiles calculated on domains of
    size $172\pi\times172\pi/\sqrt{3}$ (BM) and $236\pi\times236\pi/\sqrt{3}$ (EDRT). The full profiles are displayed
    below in Figs.~\ref{fig:density_profiles_BM} and \ref{fig:density_profiles_EDRT} together with the corresponding
    tilings. The data for the profiles are available at~\cite{ADR2022}.}
    \label{fig:density_and_lattices}
\end{figure*}

In Fig.~\ref{fig:density_and_lattices} we display portions of equilibrium density profiles obtained for both the BM and EDRT systems. 
Superimposed on two-thirds of the images are the corresponding tilings created by identifying all points $\bx_m$ that are maxima in the density profiles with $\rho(\bx_m)/\rho_m>c$, where $c=0.967$ (EDRT) or $c=0.953$ (BM) and where $\rho_m$ is the largest of all $\rho(\bx_m)$ values, and then joining neighbouring $\bx_m$ points with straight lines and paring excess vertices and edges. 
The tiles have been colored in just the left hand third of the figures. 
We have confirmed that these density profiles correspond to local minima of~$F$, but are not the global minima. 
The hexagonal crystal is the global minimum state and is the phase that typically forms from random initial conditions. 
The equilibrium QC density profiles are calculated using Picard iteration starting from an initial guess constructed in Fourier space by setting the amplitudes of $\hat{\rho}(\bk=\mathbf{0})$ and all the points on the two circles displayed in Fig.~\ref{fig:Spectra_and_wavevectors} to have a large value, while all others are given a small randomly chosen value. 
The domain on which we calculate $\rho(\bx)$ is rectangular, of size $L_x\times L_y$, with periodic boundary conditions. 
We choose the side lengths $L_x$ and $L_y$ so that the resulting (now periodic) profile is a good approximant for the true QC, with the values used chosen following the approach of Refs.~\cite{Rucklidge2009,Iooss2022}. 
Further details are given on this below in Sec.~\ref{sec:5}.

Note that in Fig.~\ref{fig:density_and_lattices} not all density peaks correspond to corners of the superimposed tiles, and in fact the tiles all contain multiple density peaks, as is common in this approach to tiling density fields.
Moreover, it is interesting to note that the manner in which the peaks decorate the tiles varies quantitatively between instances. 
The total average particle densities in these two systems are ${\bar\rho}=9.6$ (BM) and ${\bar \rho}=9.4$ (EDRT). 
Since the radii of the particles $R\approx2\pi$, these densities correspond to each particle overlapping with ${\bar\rho}\pi R^2\sim10^3$ other particles on average, which justifies the use of the mean-field approximation for $F_{ex}[\rho]$ in Eq.~\eqref{eq:F_ex}. 
Note also that these high densities make particle-based simulations of these systems difficult, requiring simulations of millions of particles if systems of the size shown in Fig.~\ref{fig:density_and_lattices} are to be achieved.

\begin{figure}
    \centering
    \includegraphics[width=0.90\hsize]{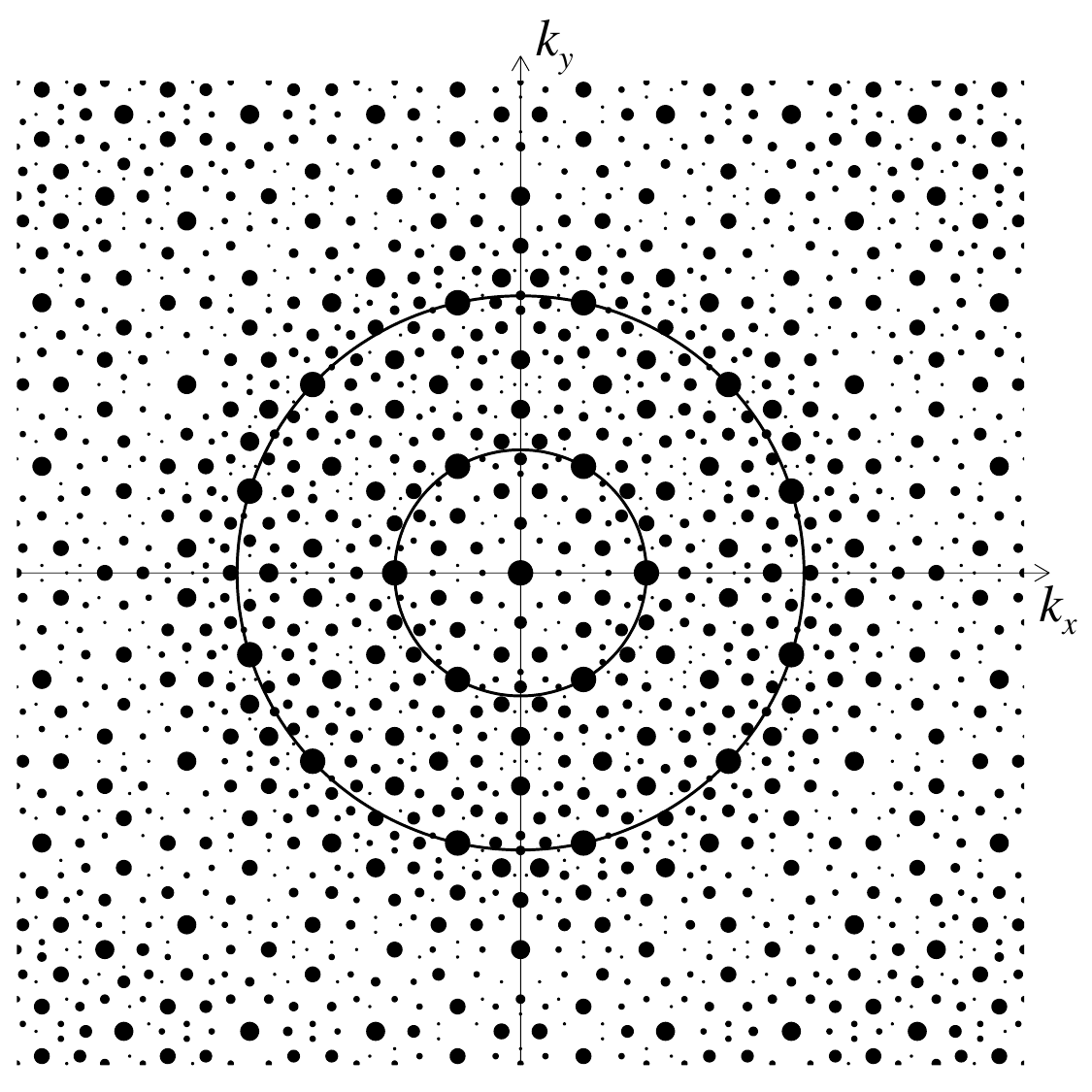}
    
    \includegraphics[width=0.90\hsize]{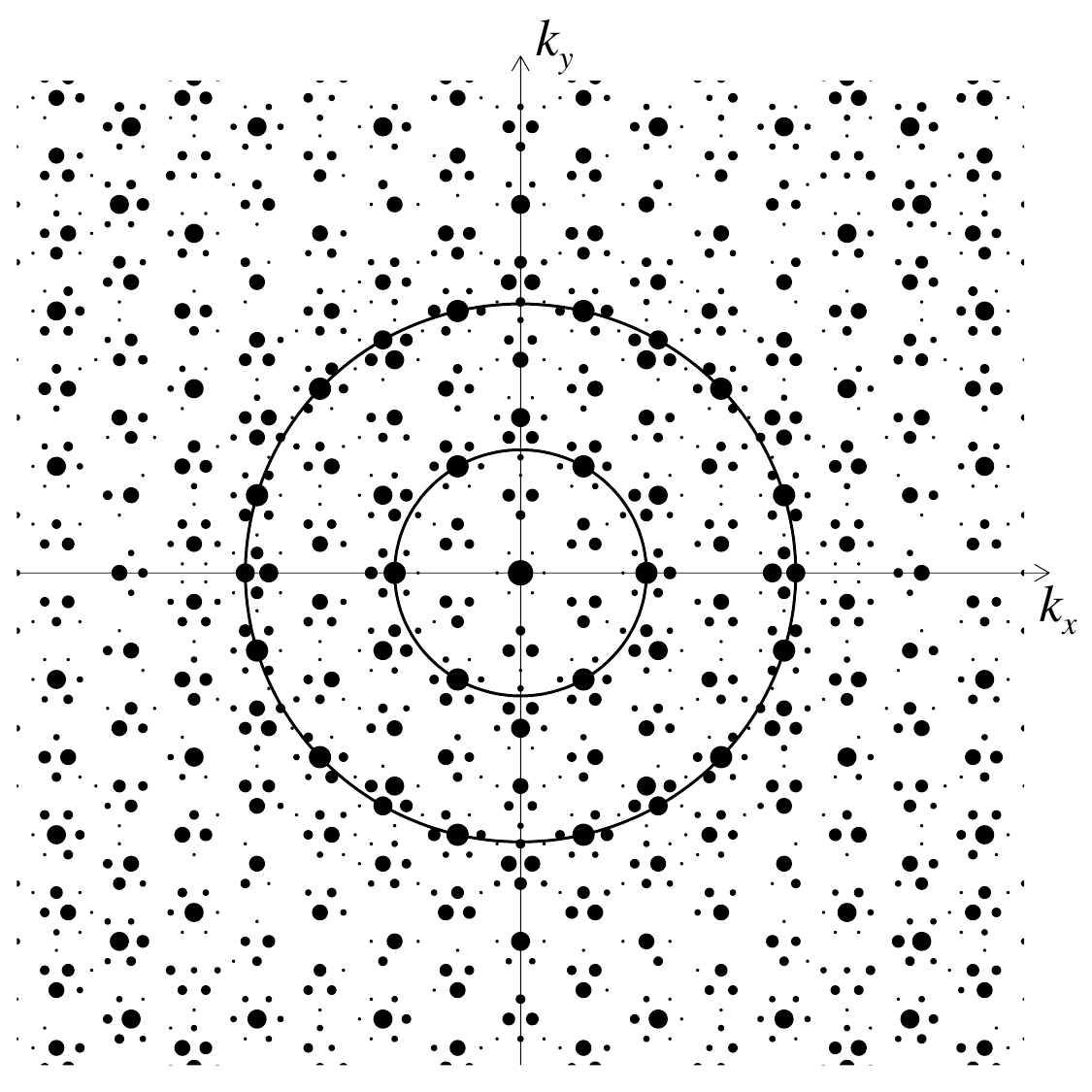}
    \caption{Power spectra (i.e.\ Fourier transforms) of the density profiles displayed in Fig.~\ref{fig:density_and_lattices}. The upper plot is for the BM system and the lower for the \hbox{EDRT}. The radius of each dot is set by the strength of the peak, with the radius going a step smaller for every factor of 10 decrease in the strength of the peak in the Fourier spectrum. The inner circle in each case is $k=1$, and the outer circles are $k=2.252$ (BM) and $k=2.186$ (EDRT). The Fourier mode amplitudes continue at reasonably large amplitude out to about $k=10$, though the equivalent power spectra for $\ln\rho$ drop off at smaller wavenumbers, consistent with the arguments of~\cite{Archer2019,Subramanian2021a}.}
    \label{fig:DFT_FT}
\end{figure}

In Fig.~\ref{fig:DFT_FT} we display the Fourier transforms of the density profiles displayed in Fig.~\ref{fig:density_and_lattices}. 
In both cases, the power spectra are dense (up to the limit imposed by the periodic domain) and have six-fold rotation symmetry.
Comparing these to the power spectra in Fig.~\ref{fig:Spectra_and_wavevectors}, and allowing for the fact that the density profiles are periodic approximants to the quasicrystals, we see that the primary peaks are the same, i.e.\ for the BM case, the six peaks on the inner circle and twelve on the outer are the same.
Similarly, for the EDRT case, the six on the inner circle and eighteen on the outer are the same. 
In terms of the locations of the peaks off the two circles, the corresponding tiling and DDFT Fourier spectra are the same, though the strength of the peaks differ. 
In particular, the DDFT spectra have far more easily visible peaks outside of the outer circle, due to the small-scale structures decorating the tiles in the DDFT system, which are not present in the tilings.

\section{Periodic approximants}
\label{sec:5}

A fundamental property of QCs is that they are not formed from any periodically repeating unit-cell: they have no translational ordering, though they do have rotational ordering. 
However, one must generally work with finite sized portions of such structures in most practical calculations such as those presented here. 
To do this we must then either (i)~deal with the complex boundary conditions that arise for finite-size portions of a QC or (ii)~we must construct periodic approximants to the true QC, which then have simple (periodic) boundary conditions. 
A third possible approach (iii)~is to note that QCs can be formed from projections of \emph{periodic} structures in higher dimensions \cite{Baake2013, Jiang2017}. 
The simplest two-dimensional QCs can generally by formed by projecting from four-dimensional periodic structures. 
Option~(ii) is the one we pursue here, though we discuss the four-dimensional nature of the EDRT tiling in Sec.~\ref{sec:6}.

There are several ways to construct the periodic approximants that we consider. 
Since we are taking a PFT approach, it is natural to focus on the wavevectors, and make small alterations to their orientations and/or lengths to make the resulting pattern periodic, as discussed for instance in~\cite{Rucklidge2009}. 
Since the quasicrystals have six-fold symmetry, we choose (as explained in detail below) wavevectors on a hexagonal lattice that approximate the most prominent wavevectors in the Fourier transform of the tiling.
These of course include the wavevectors illustrated in Fig.~\ref{fig:Spectra_and_wavevectors}. 

There are other ways of choosing periodic approximants of QCs. The cut-and-project method from the four-dimensional space~\cite{Goldman1993} works well with some classes of QCs, but in this case, there are difficulties because of the fractal nature of the projection windows (see Fig.~\ref{fig:projection_windows} below).
A third approach is to take a hexagonal arrangement of the small or large triangular tiles and then inflate these a finite number of times. 
This leads to a sequence of periodic approximants, which become better and better approximations to the true QC as the number of inflations is increased.
Below we discuss briefly the connections between this and the wavevector approaches, after first giving further details on each.

\begin{figure}[t]
\begin{center}
\includegraphics{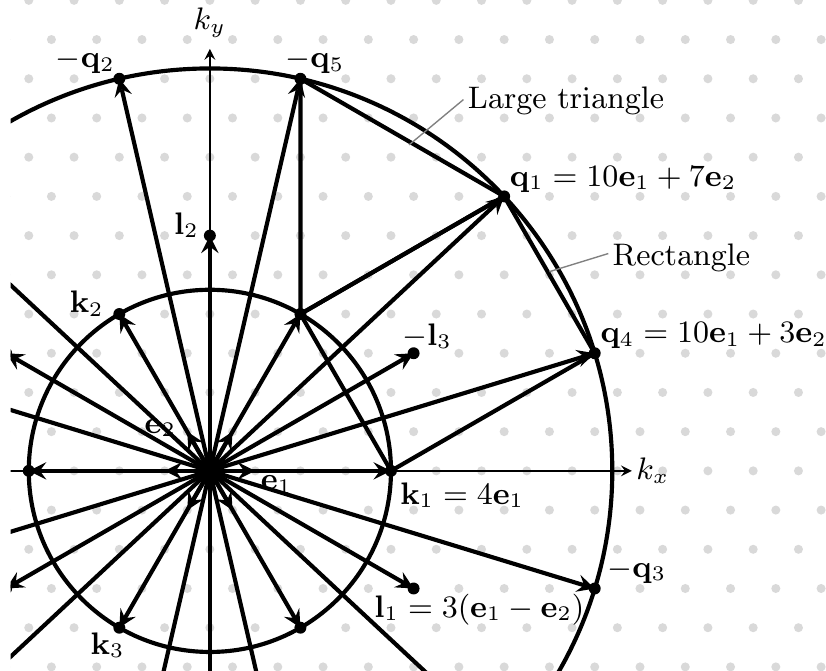}
\end{center}
\caption{The periodic hexagonal lattice (gray dots) is generated by two small vectors $\be_1$ and $\be_2$. From combinations of this pair, two circles of wavevectors of radii $k_1=1$ and~$k_2$ are built. 
The outer circle has twelve vectors: $\pm\bq_1,\pm\bq_2,\pm\bq_3,\pm\bq_4,\pm\bq_5$ and $\pm\bq_6$, and the inner has six: $\pm\bk_1,\pm\bk_2$ and $\pm\bk_3$. 
Note that these can be arranged to form rectangles (e.g.\ the one illustrated with the vectors $\pm(\bq_1+\bk_3)$ and $\pm\bk_2$ forming the sides), whose side-lengths are (in the limit) in the same ratio as the sides of the rectangles in the tiling. 
Similarly, corresponding small and large triangles can be formed. 
This case illustrated is $(a,b)=(10,7)$ and $(n,m)=(3,4)$, and the angle between $\bq_1$ and $\bq_4$ is~$\alpha\approx26^{\circ}$.}
\label{fig:hexagonal_lattice}
\end{figure}

\subsection{Constructing QC approximants in Fourier space}

We begin by defining three vectors $\be_1$, $\be_2$ and $\be_3$ of equal length and at $120^\circ$ to each other, with $\be_1$ aligned in the positive $x$~direction and
\begin{equation}
\be_1+\be_2+\be_3=0.
\end{equation}
Integer combinations of these three vectors define a hexagonal lattice -- these are the gray dots in Fig.~\ref{fig:hexagonal_lattice}.

Then, all continuous functions of interest, such as the density profile $\rho(\bx)$, can be approximated as Fourier sums of waves with wavevectors on this lattice. 
Forming them this way, the functions are periodic in a rectangular domain of size $L_x\times L_y$, where $L_x=4\pi/|\be_1|$ and $L_y=L_x/\sqrt{3}$. 

For our purposes, we follow the approach of~\cite{Dionne1997} and construct periodic approximants to BM and EDRT quasipatterns by making appropriate choices of pairs of integers $(a,b)$, with $a$ and $b$ being coprime and with $a>b>\frac{1}{2}a>0$. 
We describe below how we choose~$(a,b)$. 
The integers define six vectors~\cite{Dionne1997, Iooss2022}:
 \begin{align}
      \bq_1 &= a\be_1 + b\be_2, \nonumber\\
     \bq_2 &= (b-a)\be_1 - a\be_2, \nonumber \\
     \bq_3 &= -b\be_1 + (a-b)\be_2, \nonumber \\
     \bq_4 &= a\be_1 + (a-b)\be_2, \nonumber \\
     \bq_5 &= -b\be_1 - a\be_2, \nonumber \\
     \bq_6 &= (b-a)\be_1 + b\be_2.
 \end{align}
With this definition, we have
\begin{align}
     \bq_1+\bq_2+\bq_3=0, \nonumber\\
     \bq_4+\bq_5+\bq_6=0,
 \end{align}
 and
 \begin{align}
     |\bq_j|^2=(a^2-ab+b^2)|\be_1|^2.
 \end{align}
The twelve vectors $\pm\bq_j$ correspond to the twelve peaks on the outer circles in Fig.~\ref{fig:Spectra_and_wavevectors}. These are illustrated in Fig.~\ref{fig:hexagonal_lattice} for the case when $(a,b)=(10,7)$. The angle between $\bq_1$ and $\bq_4$ is~$\alpha$, with 
 \begin{equation}
     \cos\alpha = \frac{a^2 + 2ab - 2b^2}{2(a^2-ab+b^2)},
 \end{equation}
and
 \begin{equation}
     \sqrt{3}\sin\alpha = \frac{3a(2b-a)}{2(a^2-ab+b^2)},
 \end{equation}
noting that these are both rational numbers~\cite{Iooss2022}. 

Then we choose two sets of three vectors:
 \begin{align}
     \bk_1 &= \bq_2 - \bq_5 = (2b-a)\be_1 = m \be_1, \nonumber\\
     \bk_2 &= \bq_1 - \bq_4 = (2b-a)\be_2 = m \be_2, \nonumber\\
     \bk_3 &= \bq_3 - \bq_6 = (2b-a)\be_3 = m \be_3
 \end{align}
and
 \begin{align}
     \bl_1 &= \bq_1 + \bq_5 = (a-b)(\be_1-\be_2) = n (\be_1-\be_2), \nonumber \\
     \bl_2 &= \bq_3 + \bq_4 = (a-b)(\be_2-\be_3) = n (\be_2-\be_3), \nonumber \\
     \bl_3 &= \bq_2 + \bq_6 = (a-b)(\be_3-\be_1) = n (\be_3-\be_1).
 \end{align}
With these definitions, $|\bk_j|=(2b-a)|\be_1|$ and $|\bl_j|=\sqrt{3}(a-b)|\be_1|$. We also define $m=2b-a>0$ and $n=a-b>0$, so $a=m+2n$ and $b=m+n$.

These $\pm\bk_j$ and $\pm\bl_j$ vectors correspond to the peaks on the inner circle and just outside the inner circle in Fig.~\ref{fig:Spectra_and_wavevectors}. 
The $\bk_j$ and $\bq_j$ vectors define the triangles noted in Fig.~\ref{fig:Spectra_and_wavevectors}(b) and~\ref{fig:Spectra_and_wavevectors}(d). 
We scale $|\be_j|=1/(2b-a)=1/m$ so that $|\bk_j|=k_1=1$ and $|\bq_j|=k_2=\sqrt{a^2-ab+b^2}/(2b-a)$. 
This implies that the domain edge lengths are $L_x=2m\times2\pi$ and $L_y=L_x/\sqrt{3}$. 

This choice of scaling means that we have chosen the ratio of the two wavenumbers to be greater than two. Other choices of wavenumber ratios are possible, but we note that had we chosen the ratio to be less than two, other mode interactions would compete with those that stabilize our chosen quasicrystal~\cite{Rucklidge2012}.

Next, we show how to choose $(a,b)$ so that solving the DFT \eqref{eq:F} and/or the DDFT \eqref{eq:DDFT} in the periodic domain of size $L_x\times L_y$ will result in good approximations to the BM and EDRT quasicrystals. The aspect ratios of the rectangles in the BM tiling are:
 \begin{equation}\label{eq:aspect_ratios}
     \sqrt{3}\times\frac{1+\sqrt{13}}{6}
\end{equation}
and for the EDRT:
\begin{equation}\label{eq:aspect_ratios2}
\sqrt{3}\times\frac{3+\sqrt{33}}{12}.
 \end{equation}
Rectangles with the same aspect ratio appear in the Fourier spectrum. 
This is explained for the BM case in Ref.~\cite{Dotera2017}, but also applies to the EDRT case -- see Sec.~\ref{sec:6} below on properties of the EDRT tiling. 
The rectangle is illustrated in Fig.~\ref{fig:hexagonal_lattice}, and connects the ends of the vectors $\bk_1$, $\bq_4$, $\bq_1$ and $-\bk_3$, with short side of length $k_1=1$. 
The long side is of length $|\bl_3|=\sqrt{3}(a-b)/(2b-a)=\sqrt{3}\times\frac{n}{m}$.
So, cancelling a factor of $\sqrt{3}$ from this expression and from the irrational expressions in Eq.~\eqref{eq:aspect_ratios} leads us to consider the continued fraction approximations. 
For the BM this is:
 \begin{equation}
     \frac{n}{m} = \frac{3}{4}, \,
                   \frac{10}{13}, \,
                   \frac{33}{43}, \,
                   \frac{109}{142}, \,
                   \dots \rightarrow \frac{1+\sqrt{13}}{6}\approx0.7676
 \end{equation}
and for the EDRT:
 \begin{equation}
     \frac{n}{m} = \frac{3}{4}, \,
                   \frac{8}{11}, \,
                   \frac{43}{59}, \,
                   \frac{94}{129}, \,
                   \dots \rightarrow \frac{3+\sqrt{33}}{12}\approx0.7287.
 \end{equation}
These continued fraction approximants are readily calculated using the Euclidean algorithm and are within $\mathcal{O}(m^{-2})$ of the corresponding irrational number. 
In continued fraction notation, we have $\frac{1}{6}(1+\sqrt{13})=[0;1,3,3,3,\dots]$ and $\frac{1}{12}(3+\sqrt{33})=[0;1,2,1,2,5,2,1,2,5,\dots]$.


These choices of $(n,m)$, and the associated values of $(a,b)=(m+2n,m+n)$, define a series of periodic domain sizes (with $L_x=4\pi m$) that allow good approximations to the aperiodic BM and EDRT quasicrystals. 
For the DDFT results presented in Figs.~\ref{fig:density_and_lattices}, \ref{fig:density_profiles_BM} and~\ref{fig:density_profiles_EDRT}, we used $(n,m)=(33,43)$ (BM case) and $(n,m)=(43,59)$ (EDRT case), in domains with $L_x=86\times2\pi$ and $L_x=118\times2\pi$, respectively. 
These provide rational approximations to the irrational rectangle tile aspect ratio that are within $0.02\%$ of the true value. For larger~$(n,m)$, the error goes as~$m^{-2}$.

The method presented here of constructing periodic approximants to six-fold quasicrystals generalizes the method proposed in Ref.~\cite{Rucklidge2009} for the dodecagonal quasicrystal. 
That method was based on square periodic domains and a different series of rational approximations to an irrational number. 
Here the $1:\sqrt{3}$ domains lend themselves naturally to quasicrystals with six-fold symmetry, and with this point of view, the relevant irrational number for the dodecagonal case is $1/\sqrt{3}$. 
The periodic approximants have $\frac{n}{m}=\frac{3}{5},\frac{4}{7},\frac{11}{19},\frac{15}{26},\frac{41}{71},\frac{56}{97},\dots\rightarrow\frac{1}{\sqrt{3}}$.


\begin{figure*}[t]
\begin{center}
\includegraphics{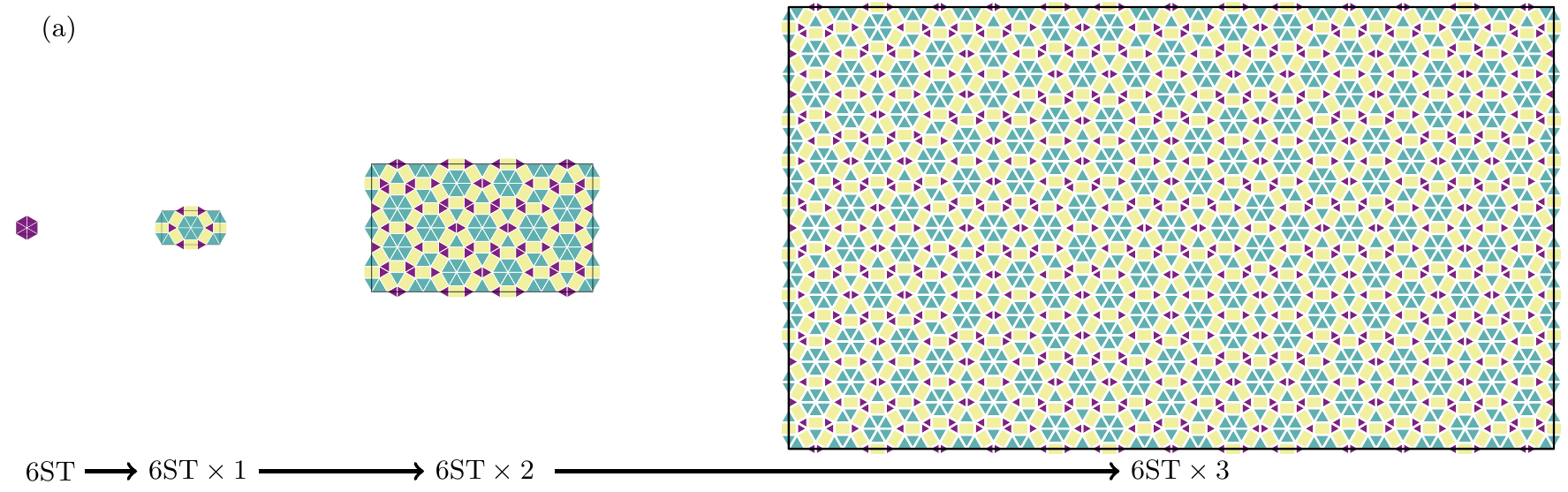}

\vspace{1ex}

\includegraphics{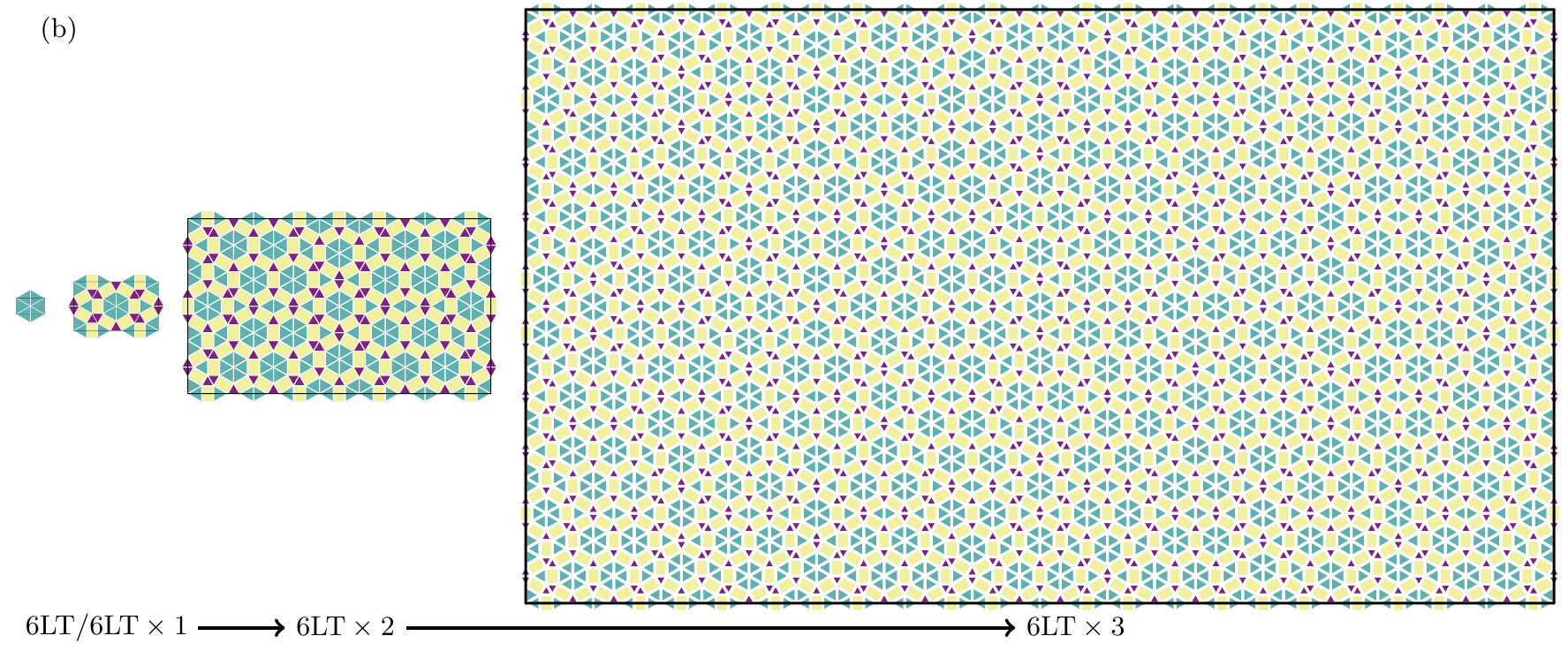}
\end{center}
\caption{\label{fig:BM_rectangular_approximants}
Constructing BM QC approximants by inflation: (a) six small triangles (ST) or (b)~six large triangles (LT) are arranged to form a hexagon, with a periodic rectangle with aspect ratio $\sqrt{3}:1$ indicated in black (allowing the tiles to overlap). The six triangles are then inflated three times, using the inflation rules in Fig.~\ref{fig:BM_EDRT_inflation}. The scaling is consistent within the panels, and tiles that are completely outside the periodic rectangle are not drawn. The last inflated tiling in~(a) is repeated in Fig.~\ref{fig:density_profiles_BM}, which compares it to the DDFT density profile.}
\end{figure*}

\begin{figure*}[t]
\begin{center}
\includegraphics{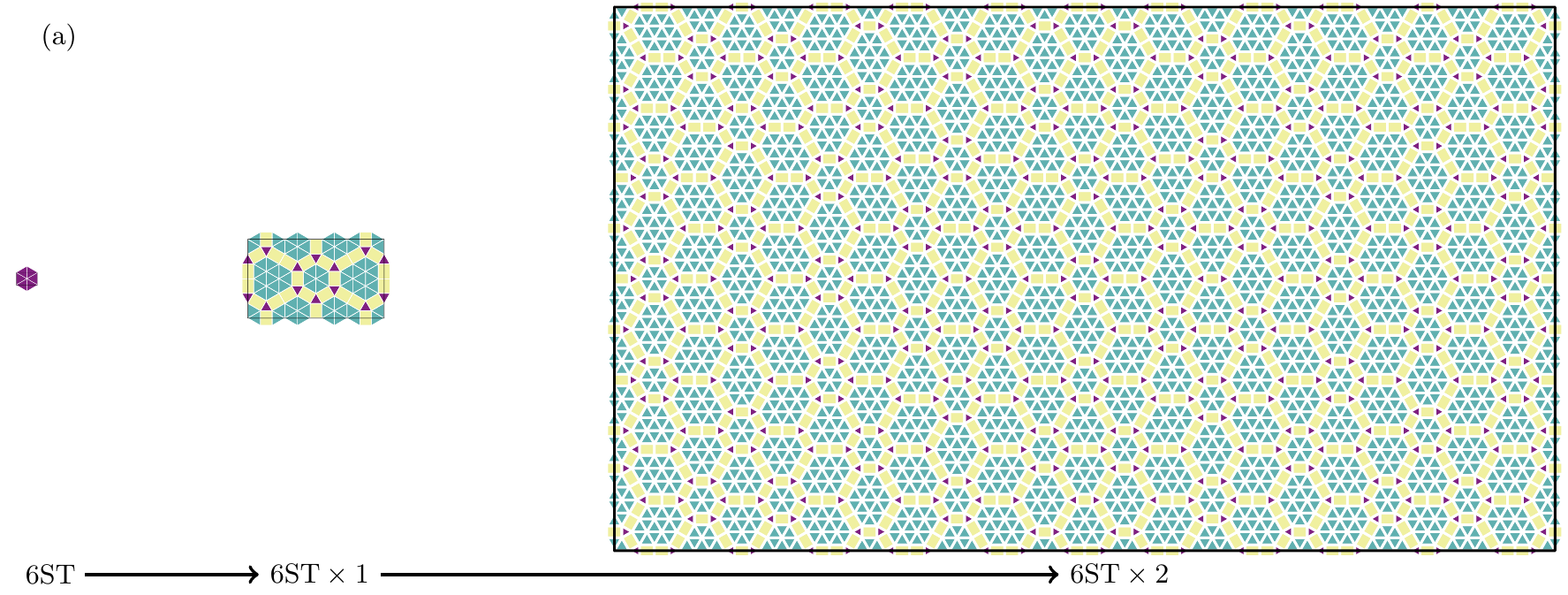}

\vspace{1ex}

\includegraphics{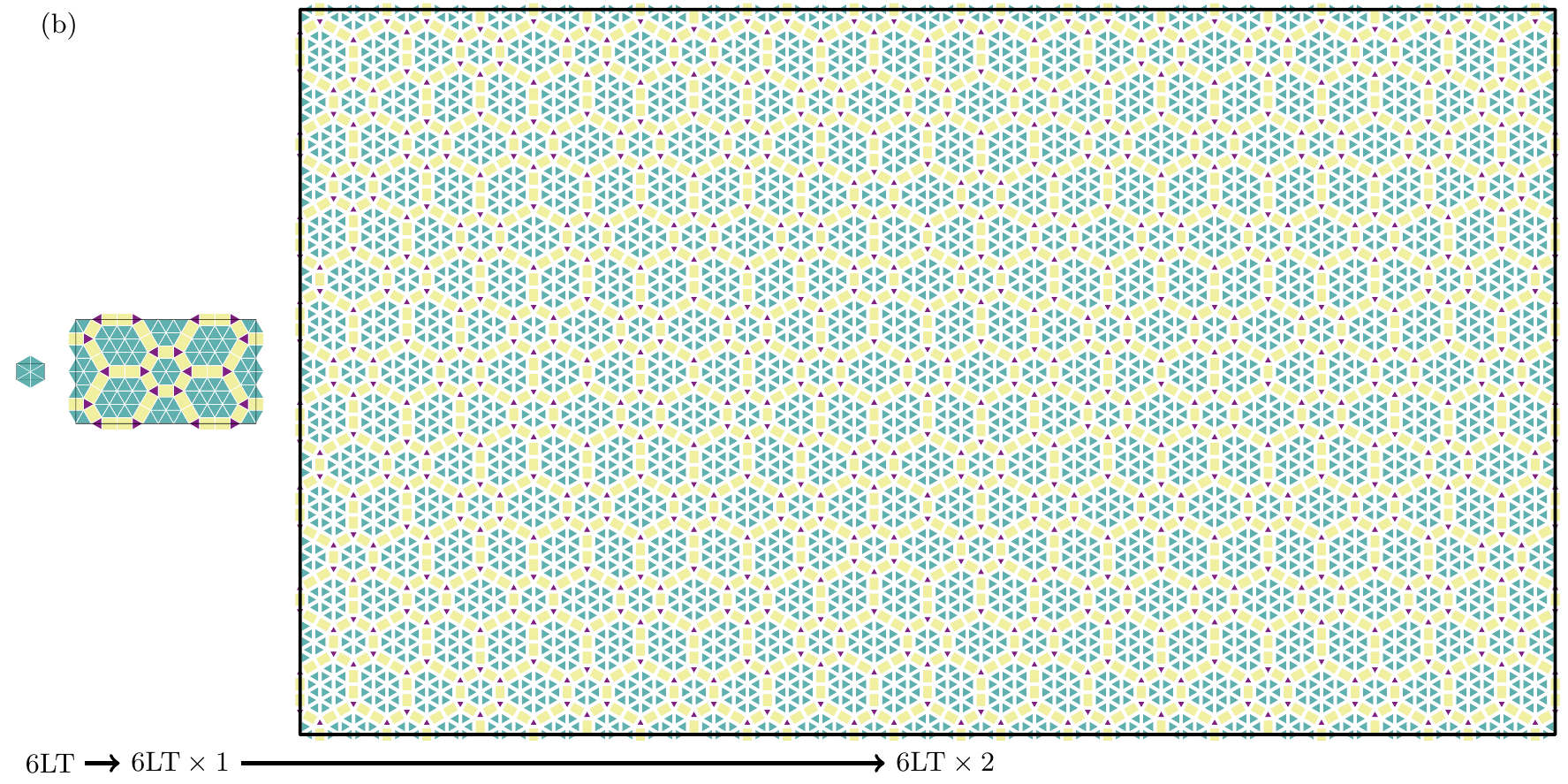}
\end{center}
\caption{\label{fig:EDRT_rectangular_approximants}
Constructing EDRT QC approximants by inflation: (a)~six small or (b)~large triangles are arranged to form a hexagon, as in Fig.~\ref{fig:BM_rectangular_approximants}. The six triangles are then inflated twice. The last inflated tiling in~(a) is repeated in Fig.~\ref{fig:density_profiles_EDRT}, which compares it to the DDFT density profile.}
\end{figure*}

\subsection{Constructing QC approximants via inflation}

As mentioned above, an alternative method of constructing periodic approximants is based on tiles and inflation rules, rather than wavevectors. 
There are several ways to construct a sequence of approximants that tends to ideal QCs. 
Here we illustrate a series of rectangular approximants with aspect ratio $\sqrt{3}:1$, natural for structures with six-fold symmetry. 
The starting points are either six large or six small triangles, arranged as shown in Figs.~\ref{fig:BM_rectangular_approximants} and~\ref{fig:EDRT_rectangular_approximants}.
With periodic boundary conditions, these form a rectangle with aspect ratio $\sqrt{3}:1$. 
On applying the inflation rules, in which the number of polygons grows as described by Eq.~\eqref{eq:SLTR_inflation} below, we obtain a sequence of periodic approximants of increasing size. 
Ideally, the starting choice of tiles should be a configuration that would appear naturally within the full tiling: this is the case when choosing six large triangles but not when choosing six small triangles.

\subsection{Linking the Fourier and tiling viewpoints}

\begin{figure*}
    \centering
    \includegraphics[width=0.91\linewidth]{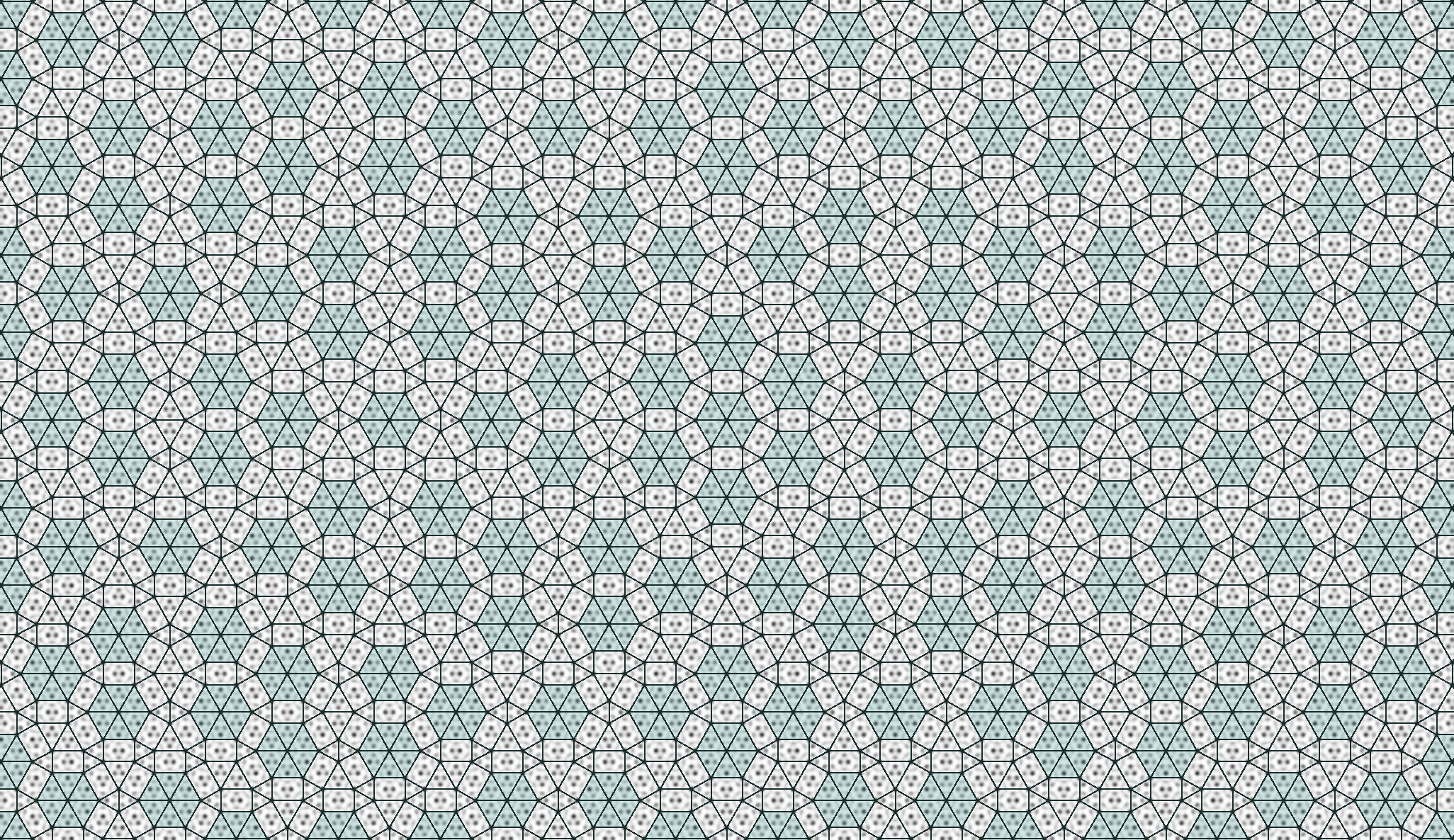}

    \includegraphics[width=0.93\linewidth]{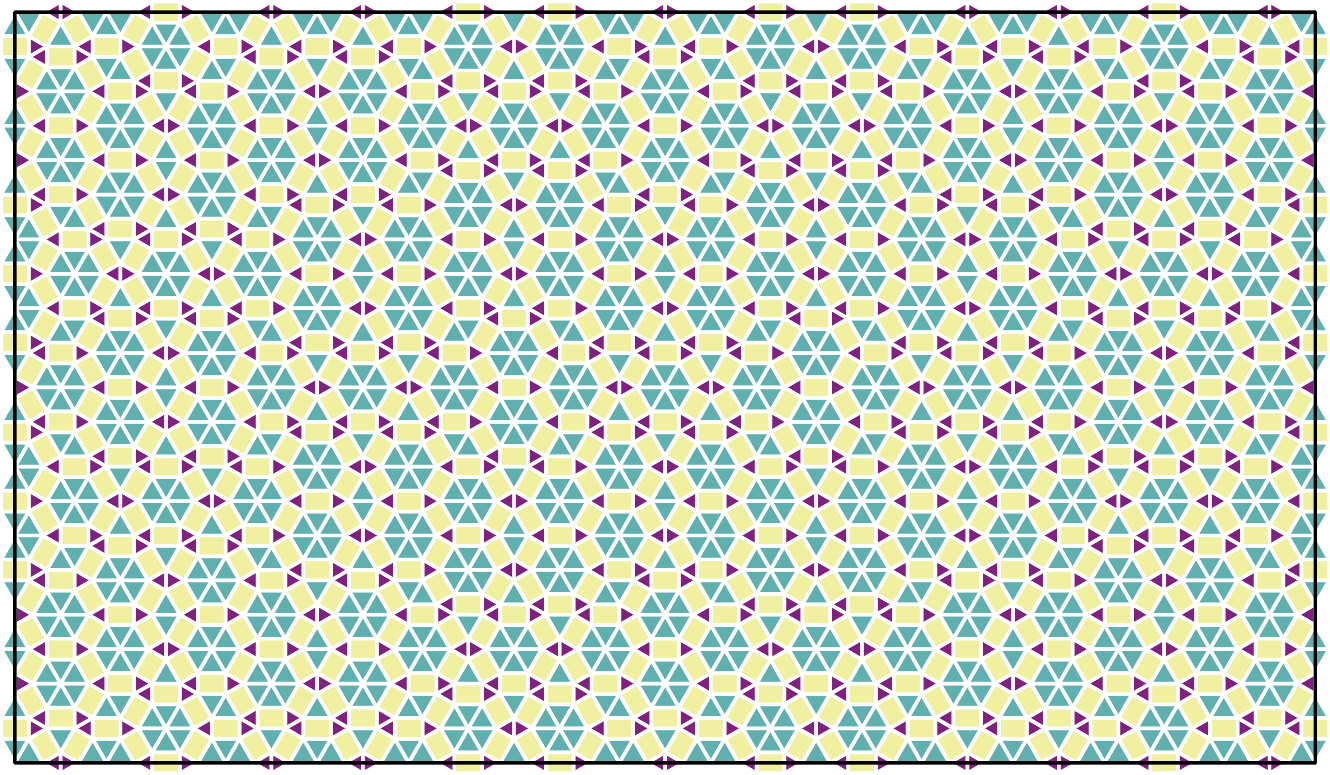}
    \caption{Density profiles in the BM case with the choice $(m,n)=(43,33)$ and the tiling corresponding to six small
    triangles inflated three times (see Fig.~\ref{fig:BM_rectangular_approximants}). Sets of six large triangles in
    the density field are tinted green to guide the eye: the two tilings differ in their central and corner regions,
    but small rearrangements of the choice of tiles (phason flips) in the density field would lead to exact
    agreement. The data for the density profile are available at~\cite{ADR2022}.}

    \label{fig:density_profiles_BM}
\end{figure*}

\begin{figure*}
    \centering
    \includegraphics[width=0.91\linewidth]{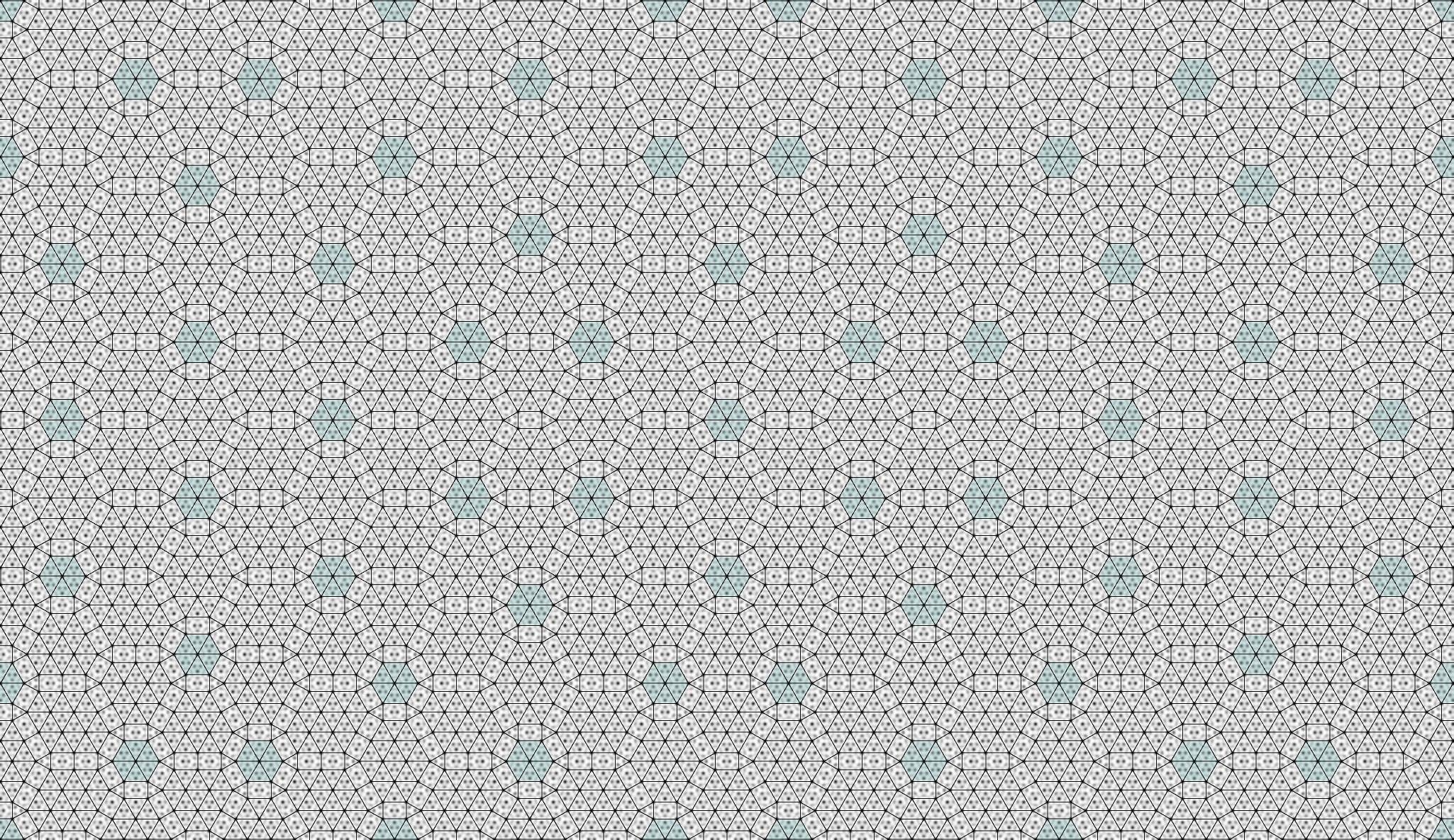}

    \includegraphics[width=0.925\linewidth]{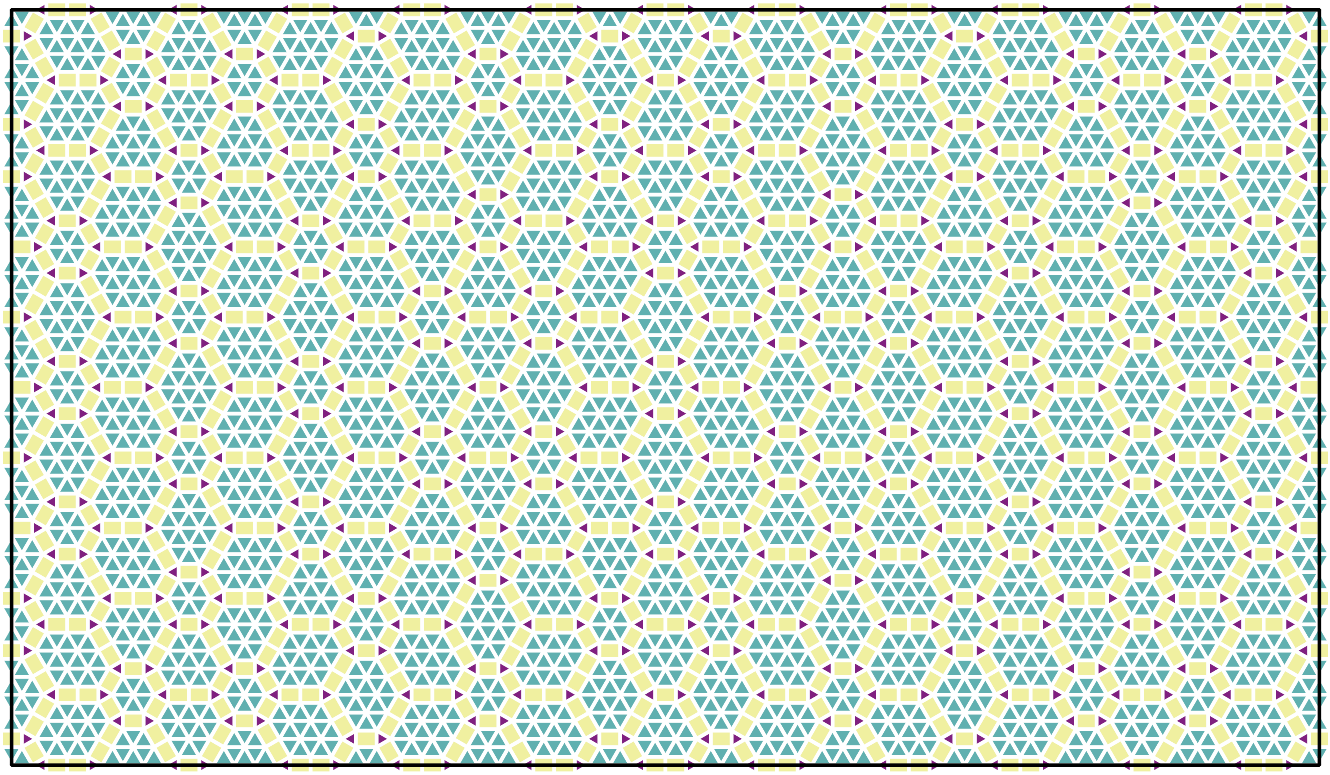}
    \caption{Density profiles in the EDRT case with the choice $(m,n)=(59,43)$ and the tiling corresponding to six
    small triangles inflated twice (see Fig.~\ref{fig:EDRT_rectangular_approximants}). Sets of six large triangles in
    the density field are tinted green to guide the eye: the two tilings are the same. The data for the density
    profile are available at~\cite{ADR2022}.}
    \label{fig:density_profiles_EDRT}
\end{figure*}

In Figs.~\ref{fig:density_profiles_BM} and \ref{fig:density_profiles_EDRT} we show the full extent of the DDFT density profiles from which portions are displayed in Fig.~\ref{fig:density_and_lattices}. 
Figure~\ref{fig:density_profiles_BM} shows the BM QC and Fig.~\ref{fig:density_profiles_EDRT} shows the EDRT \hbox{QC}. 
In each case we superimpose the corresponding tiling with vertices at a sub-set of the maxima in the density profile, as described above in Sec.~\ref{sec:4}. 
We tint in green some of the large triangles to aid the eye. 
Beneath each density profile, we display the corresponding tiling approximants, which are also the last in the sequence of inflations illustrated in Fig.~\ref{fig:BM_rectangular_approximants}(a) and Fig.~\ref{fig:EDRT_rectangular_approximants}(a).

Notice in Fig.~\ref{fig:density_profiles_BM} that arrangements of six large triangles (highlighted in green) in central and corner parts of the density profile do not line up with the corresponding sets of six large triangles in the tiling. 
The reason for this is that the rules we use for linking maxima in the density profile are local and do not take into account the global structure imposed by the inflation rules.
Minor rearrangements (\emph{phason flips}) of the tiles would make the match perfect. 
In contrast, in the EDRT case (Fig.~\ref{fig:density_profiles_EDRT}), the agreement is perfect without any rearrangement. 
Other choices of $(m,n)$ in the sequence of approximations in the DDFT calculations lead, in both the BM and EDRT cases, to tilings that match those that can be found by sequences of inflations of six small or large triangles (though the exact alignment between these is more complicated in the EDRT case).

\section{Properties of the EDRT tiling}
\label{sec:6}

The EDRT tiling presented in Figs.~\ref{fig:BM_EDRT_inflation}, \ref{fig:BM_EDRT_R_example} and \ref{fig:EDRT_rectangular_approximants} is composed of small (ST) and large (LT) equilateral triangles of edge lengths $S$ and $L$, respectively, and of rectangles~(R) of size  $L\times S$; see Fig.~\ref{fig:BM_EDRT_inflation}. We impose the equidiagonal conditions for ST--R, LT--R and LT--LT pairs as shown in Fig.~\ref{fig:BM_EDRT_inflation}(h). To do this, the ratio of the edge lengths must be
\begin{equation}
\phi=\frac{L}{S}=\frac{\sqrt{3}+\sqrt{11}}{4}\approx1.262.
\label{eq:phi_ratio}
\end{equation}
In each inflation step, the tiles are subdivided according to the rules illustrated in Fig.~\ref{fig:BM_EDRT_inflation}(e)--(g). Upon subdivision, the two lengths of the $i$-th generation tiling $L_i$ and $S_i$ transform as
\begin{equation}
\left(\begin{array}{c}
		L_{i+1}\\
		S_{i+1}\\
	\end{array}
\right)
= \left(\begin{array}{ccc}
		 3\sqrt{3} &2 &  \\
		 4 & \sqrt{3} \\
		\end{array}
\right)
\left(\begin{array}{c}
		L_{i}\\
		S_{i}\\
	\end{array}
\right).
\label{eq:transformationrule}
\end{equation}
The positive eigenvalue of this transformation matrix is
\begin{equation}
\beta=2\sqrt{3}+\sqrt{11}\approx 6.781,
\end{equation}
which is the inflation factor, and the corresponding eigenvector gives exactly the ratio in Eq.~\eqref{eq:phi_ratio}.
Note that the inflation factor $\beta$ is not a Pisot number but its square $\beta^2=23+4\sqrt{33}$, corresponding to two consecutive subdivisions, is a Pisot number.
We note that the EDRT tiling is categorized as a type~IIC tiling, extending the scheme of Ref.~\cite{Nakakura2019}. More precisely type~II means the tiles are rotated by $30^{\circ}$ at each inflation, and has in this case $n=3$ and $m=4$ in Eq.~(16) in the main text of Ref.~\cite{Nakakura2019} (compare with Eq.~\eqref{eq:transformationrule} above),  and the label~C is given in analogy to the case of type~IC in Supplementary Note~3 of that paper.

Also of interest are the numbers of long and short edges $n^L_i$ and $n^S_i$, respectively, which transform according to
\begin{equation}
\left(\begin{array}{c}
		n^L_{i+1}\\
		n^S_{i+1}\\
	\end{array}
\right)
= \left(\begin{array}{ccc}
		 3\sqrt{3} &4 &  \\
		 2 & \sqrt{3} \\
		\end{array}
\right)
\left(\begin{array}{c}
		n^L_{i}\\
		n^S_{i}\\
	\end{array}
\right).
\label{eq:transformationrule2}
\end{equation}
In a self-similar tiling, the ratio of the numbers of long and short edges is 
\begin{equation}
\psi=\frac{\sqrt{3}+\sqrt{11}}{2}\approx 2.524,
\end{equation}
coming from an eigenvector of the matrix in~\eqref{eq:transformationrule2}.




By inspecting the inflation rules for each tile type in Fig.~\ref{fig:BM_EDRT_inflation}(e)--(g), we find that in the EDRT tiling the numbers of ST, LT, and R tiles in the $(i+1)$-th generation denoted by $ST_{i+1}$, $LT_{i+1}$, and $R_{i+1}$, respectively, are related to those in the $i$-th generation by 
\begin{equation}
\left(\!\begin{array}{c}
		ST_{i+1}\\
		LT_{i+1}\\
		R_{i+1}\\
	\end{array}\!\!
\right)
=\left(
\begin{array}{ccc}
3 & 4  & 8 \\
16 & 27 & 48 \\
6 & 9 & 17 \\
\end{array}
\right)\!\!
\left(\!\begin{array}{c}
		ST_{i}\\
		LT_{i}\\
		R_{i}\\
	\end{array}\!
\right).
\label{eq:SLTR_inflation}
\end{equation}
The largest eigenvalue of the above matrix is $\beta^2$. 
The eigenvector corresponding to $\beta^2$ is 
 \begin{equation}
 \frac{1}{12}
 \left(\!\begin{array}{c}
		7-\sqrt{33}\\
		8\\
		-3+\sqrt{33}\\
	\end{array}\!\right)
 \approx
 \left(\!\begin{array}{c}
		0.105\\
		0.667\\
		0.229\\
	\end{array}\!\right),
 \end{equation}
while for the BM tiling the corresponding vector is \cite{Dotera2017}:
 \begin{equation}
 \frac{1}{43}
 \left(\!\begin{array}{c}
		29-5\sqrt{13}\\
		23-\sqrt{13}\\
		-9+6\sqrt{13}\\
	\end{array}\!\right)
 \approx
 \left(\!\begin{array}{c}
		0.255\\
		0.451\\
		0.294\\
	\end{array}\!\right).
 \end{equation}
Therefore, we find that the EDRT tiling has a higher proportion of the~\hbox{LT} compared to the BM tiling. 


\begin{figure}[t]
\begin{center}
\includegraphics{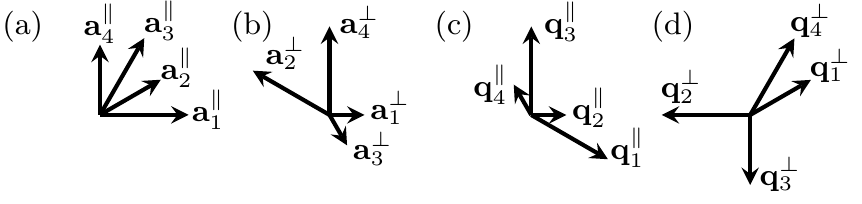}
\end{center}
\vspace{-6ex}
\caption{\label{vectors} Basis vectors in real and reciprocal space for the EDRT tiling.
Projected basis vectors in the physical and in the perpendicular space, ${\ba}^{\parallel}_j$ and ${\ba}^{\perp}_j$ are shown in (a) and (b), respectively.
Projected reciprocal-space basis vectors in the physical and in the perpendicular space,
${\bq}^{\parallel}_j$ and ${\bq}^{\perp}_j$ are shown in (c) and (d), respectively.}
\end{figure}


\begin{figure}[t]
\begin{center}
\includegraphics{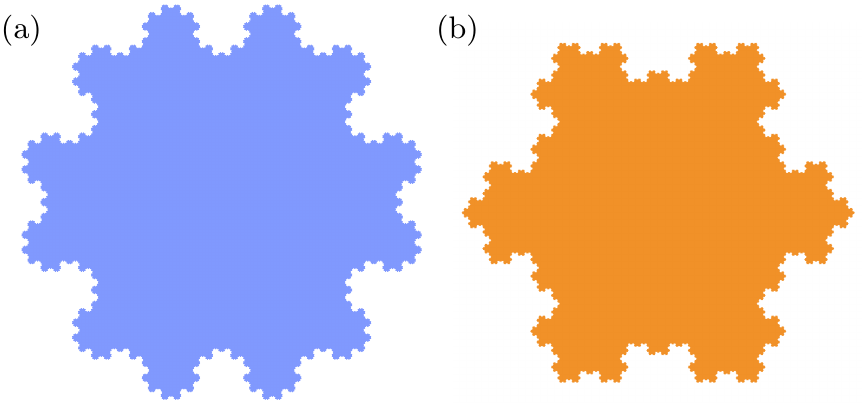}
\end{center}
\caption{\label{fig:projection_windows}
Projection windows for (a)~the EDRT tiling with 136525 points and (b)~the BM tiling.}
\end{figure}

The two-dimensional aperiodic EDRT tiling can be constructed by projecting from a four-dimensional periodic structure, with set of basis vectors $\{{\bm a}_j\}$, where $j=0,1,2,3$. Projection matrices $P^{\parallel}$ and $P^{\perp}$ allow us to define the basis vectors in the physical and in the perpendicular space by ${\bm a}^{\parallel}_j \equiv P^{\parallel} {\bm a}_j$ and ${\bm a}^{\perp}_j \equiv P^{\perp} {\bm a}_j$, respectively. See the section ``Higher-dimensional representation'' in Ref.~\cite{Dotera2017} for more details. Thus, we find that
$\left|{\bm a}_{\rm odd}^{\parallel}\right|=a\, \ell\,\alpha$, $\left|{\bm a}_{\rm even}^{\parallel}\right|=c\, \alpha$, $\left|{\bm a}_{\rm odd}^{\perp}\right|=a\, \alpha$, and $\left|{\bm a}_{\rm even}^{\perp}\right|=c\, \ell\,\alpha$. The ratios of the lengths of even and odd basis vectors in the physical and in the perpendicular space are
\begin{equation}
\frac{\left|{\bm a}_{\rm odd}^{\parallel}\right|}{\left|{\bm a}_{\rm even}^{\parallel}\right|}
=\frac{a\ell}{c}=\phi
\end{equation}
and
\begin{equation}
\frac{\left|{\bm a}_{\rm even}^{\perp}\right|}{\left|{\bm a}_{\rm odd}^{\perp}\right|}
=\frac{c\ell}{a}=\psi,
\end{equation}
respectively.

\begin{table}[t]
\begin{center}
\begin{tabular}{|c|cccc|c|c|c|}
\hline
 \ No. \ & \ $n_1$& $n_2$ & $n_3$ & $n_3$ \ & Intensity &\ $|\bm{k}^\perp|$ \ & \ $|\bm{k}^\parallel|$ \ \\
\hline 
0 & 0 & 0 & 0 & 0 & 1.000 & 0.0 & 0.0 \\
\hline
1 & 2 & 2 & 2 & 1 & 0.936 & 0.571 & 14.724  \\
\hline
2& 3 & 3 & 4 & 2 & 0.818 & 0.990 & 25.503\\
\hline
3& 2 & 3 & 4 & 3 & 0.818 & 0.990 & 25.503\\
\hline
4& 1 & 1 & 2 & 1 & 0.648 & 1.443 & 11.666\\
\hline
5& 1 & 2 & 3 & 2 & 0.603 & 1.552 & 18.786\\
\hline
6& 1 & 1 & 1 & 1 & 0.437 & 1.959 & 7.442\\
\hline
\end{tabular}
\caption{\label{indextable} List of prominent peaks for the EDRT tiling having intensity $I(\bm{k}^\parallel) > 0.4$. Note that the ratio of $|\bm{k}^\parallel|$ for the second and third intensity peaks to the fourth is $25.503/11.666=2.186$.}
\end{center}
\end{table}

Using these ratios, we construct the projection windows in the perpendicular space. The position of a vertex of a tiling in the physical space is given as ${\bm r}^\parallel=\sum_{j=0}^3 n_j {\bm a}_j^\parallel$, where $n_j$ are integers and ${\bm a}_j^\parallel$ are the physical-space basis vectors of the tiling. 
The same set of $n_i$ also defines the position of the corresponding vertex in the perpendicular space ${\bm r}^\perp=\sum_{j=0}^3 n_j {\bm a}_j ^\perp$, and these vertices constitute the projection windows. 
Figure~\ref{fig:projection_windows}(a) shows the projection window, which tends to a self-similar shape as one proceeds with inflation from generation to generation. Figure~\ref{fig:projection_windows}(b) shows the projection window for the BM tiling.

We find that
$\left|{\bm q}_{\rm odd}^{\parallel}\right|= 2\pi\ell\alpha /a$, 
$\left|{\bm q}_{\rm even}^{\parallel}\right|=2\pi\alpha/c$, 
$\left|{\bm q}_{\rm odd}^{\perp}\right|=2\pi\alpha/a$, and 
$\left|{\bm q}_{\rm even}^{\perp}\right|=2\pi\ell\alpha/c$. 
Thus the ratios of lengths of the reciprocal-space basis vectors in the physical and the perpendicular space are
\begin{equation}
\frac{\left|{\bm q}^\parallel_{\rm odd}\right|}
{\left|{\bm q}^\parallel_{\rm even}\right|}=\frac{c\ell}{a}=\psi
\end{equation}
and
\begin{equation}
\frac{\left|{\bm q}^\perp_{\rm even}\right|}
{\left|{\bm q}^\perp_{\rm odd}\right|}=\frac{a\ell}{c}=\phi,
\end{equation}
respectively. These ratios are the inverses of those in the physical space.

Recall that Fig.~\ref{fig:Spectra_and_wavevectors}(c) shows the Fourier transform of a finite but large EDRT tiling, with the intensities normalized by the central peak. Remarkably, many of the prominent peaks and numerous of the smaller ones are similar to those for the bronze-mean tiling [Fig.~\ref{fig:Spectra_and_wavevectors}(a)]. The first seven strong peaks are listed in Table~\ref{indextable}. Note that strong peaks No.~2 and 3 have the same length wavevectors, which is nothing but the equidiagonal property.

\section{Concluding remarks}
\label{sec:7}

To conclude, we recall that aperiodic tilings have been invoked as a description of the geometry of QCs ever since their discovery~\cite{Levine1984}, and having two length scales present in a system is known to stabilize QCs~\cite{Lifshitz1997,Lifshitz2007a}. Here, we have demonstrated how to join these approaches together, leading to a new example of a QC/tiling, through analysis of the Fourier spectrum of the aperiodic tiling and a careful choice of interaction potential. The interaction potential we use~\eqref{eq:v} is a model for cluster crystals of polymer micelles or dendrimers with a core and soft corona~\cite{Mladek2008, Lenz2012, Barkan2014}. Thus, our work contributes to understanding how to design soft-matter systems to form particular structures that could have useful (e.g.~optical) properties.

More elaborate potentials, perhaps involving three-body interactions, may be required for other tilings or indeed to make the structures discussed here the global minima of~$F$. 
The number of aperiodic tilings found so far is large~\cite{Frettloh2021}, and includes structures that may be relevant to two-dimensional materials such as bilayer graphene~\cite{Iooss2022,Zeller2014} and three-dimensional quasicrystals~\cite{Subramanian2021a}. We are optimistic that our approach can be used, at least in principle, to find soft-particle systems that self-assemble into these structures.

The data associated with this paper are openly available from the University of Leeds Data 
Repository~\cite{ADR2022}.

\goodbreak

\begin{acknowledgments}
We gratefully acknowledge stimulating discussions with Michael Baake, Uwe Grimm, Dominic Hopkinson, G\'erard Iooss, Ron Lifshitz, Dan Ratliff and Priya Subramanian, and to Jessica Wise for support. 
We are grateful for the kind hospitality of Kindai University, where this work was initiated. 
We acknowledge funding support from the EPSRC (EP/P015689/1, A.J.A. and EP/P015611/1, A.M.R.) and from the Japan Society for the Promotion of Science through Grant-in-Aid for Scientific Research (C) (No.~19K03777).
This paper is dedicated to the memory of Uwe Grimm.
\end{acknowledgments}



\begin{thebibliography}{53}%
\makeatletter
\providecommand \@ifxundefined [1]{%
 \@ifx{#1\undefined}
}%
\providecommand \@ifnum [1]{%
 \ifnum #1\expandafter \@firstoftwo
 \else \expandafter \@secondoftwo
 \fi
}%
\providecommand \@ifx [1]{%
 \ifx #1\expandafter \@firstoftwo
 \else \expandafter \@secondoftwo
 \fi
}%
\providecommand \natexlab [1]{#1}%
\providecommand \enquote  [1]{``#1''}%
\providecommand \bibnamefont  [1]{#1}%
\providecommand \bibfnamefont [1]{#1}%
\providecommand \citenamefont [1]{#1}%
\providecommand \href@noop [0]{\@secondoftwo}%
\providecommand \href [0]{\begingroup \@sanitize@url \@href}%
\providecommand \@href[1]{\@@startlink{#1}\@@href}%
\providecommand \@@href[1]{\endgroup#1\@@endlink}%
\providecommand \@sanitize@url [0]{\catcode `\\12\catcode `\$12\catcode
  `\&12\catcode `\#12\catcode `\^12\catcode `\_12\catcode `\%12\relax}%
\providecommand \@@startlink[1]{}%
\providecommand \@@endlink[0]{}%
\providecommand \url  [0]{\begingroup\@sanitize@url \@url }%
\providecommand \@url [1]{\endgroup\@href {#1}{\urlprefix }}%
\providecommand \urlprefix  [0]{URL }%
\providecommand \Eprint [0]{\href }%
\providecommand \doibase [0]{https://doi.org/}%
\providecommand \selectlanguage [0]{\@gobble}%
\providecommand \bibinfo  [0]{\@secondoftwo}%
\providecommand \bibfield  [0]{\@secondoftwo}%
\providecommand \translation [1]{[#1]}%
\providecommand \BibitemOpen [0]{}%
\providecommand \bibitemStop [0]{}%
\providecommand \bibitemNoStop [0]{.\EOS\space}%
\providecommand \EOS [0]{\spacefactor3000\relax}%
\providecommand \BibitemShut  [1]{\csname bibitem#1\endcsname}%
\let\auto@bib@innerbib\@empty
\bibitem [{\citenamefont {Baake}\ and\ \citenamefont
  {Grimm}(2013)}]{Baake2013}%
  \BibitemOpen
  \bibfield  {author} {\bibinfo {author} {\bibfnamefont {M.}~\bibnamefont
  {Baake}}\ and\ \bibinfo {author} {\bibfnamefont {U.}~\bibnamefont {Grimm}},\
  }\href {https://doi.org/10.1017/CBO9781139025256} {\emph {\bibinfo {title}
  {Aperiodic Order. Volume 1: A Mathematical Invitation}}}\ (\bibinfo
  {publisher} {Cambridge University Press},\ \bibinfo {year}
  {2013})\BibitemShut {NoStop}%
\bibitem [{\citenamefont {Shechtman}\ \emph {et~al.}(1984)\citenamefont
  {Shechtman}, \citenamefont {Blech}, \citenamefont {Gratias},\ and\
  \citenamefont {Cahn}}]{Shechtman1984a}%
  \BibitemOpen
  \bibfield  {author} {\bibinfo {author} {\bibfnamefont {D.}~\bibnamefont
  {Shechtman}}, \bibinfo {author} {\bibfnamefont {I.}~\bibnamefont {Blech}},
  \bibinfo {author} {\bibfnamefont {D.}~\bibnamefont {Gratias}},\ and\ \bibinfo
  {author} {\bibfnamefont {J.~W.}\ \bibnamefont {Cahn}},\ }\bibfield  {title}
  {\bibinfo {title} {Metallic phase with long-range orientational order and no
  translational symmetry},\ }\href
  {https://doi.org/10.1103/PhysRevLett.53.1951} {\bibfield  {journal} {\bibinfo
   {journal} {Phys. Rev. Lett.}\ }\textbf {\bibinfo {volume} {53}},\ \bibinfo
  {pages} {1951} (\bibinfo {year} {1984})}\BibitemShut {NoStop}%
\bibitem [{\citenamefont {Zeng}\ \emph {et~al.}(2004)\citenamefont {Zeng},
  \citenamefont {Ungar}, \citenamefont {Liu}, \citenamefont {Percec},
  \citenamefont {Dulcey},\ and\ \citenamefont {Hobbs}}]{Zeng2004}%
  \BibitemOpen
  \bibfield  {author} {\bibinfo {author} {\bibfnamefont {X.~B.}\ \bibnamefont
  {Zeng}}, \bibinfo {author} {\bibfnamefont {G.}~\bibnamefont {Ungar}},
  \bibinfo {author} {\bibfnamefont {Y.~S.}\ \bibnamefont {Liu}}, \bibinfo
  {author} {\bibfnamefont {V.}~\bibnamefont {Percec}}, \bibinfo {author}
  {\bibfnamefont {S.~E.}\ \bibnamefont {Dulcey}},\ and\ \bibinfo {author}
  {\bibfnamefont {J.~K.}\ \bibnamefont {Hobbs}},\ }\bibfield  {title} {\bibinfo
  {title} {Supramolecular dendritic liquid quasicrystals},\ }\href
  {https://doi.org/10.1038/nature02368} {\bibfield  {journal} {\bibinfo
  {journal} {Nature}\ }\textbf {\bibinfo {volume} {428}},\ \bibinfo {pages}
  {157} (\bibinfo {year} {2004})}\BibitemShut {NoStop}%
\bibitem [{\citenamefont {Hayashida}\ \emph {et~al.}(2007)\citenamefont
  {Hayashida}, \citenamefont {Dotera}, \citenamefont {Takano},\ and\
  \citenamefont {Matsushita}}]{Hayashida2007}%
  \BibitemOpen
  \bibfield  {author} {\bibinfo {author} {\bibfnamefont {K.}~\bibnamefont
  {Hayashida}}, \bibinfo {author} {\bibfnamefont {T.}~\bibnamefont {Dotera}},
  \bibinfo {author} {\bibfnamefont {A.}~\bibnamefont {Takano}},\ and\ \bibinfo
  {author} {\bibfnamefont {Y.}~\bibnamefont {Matsushita}},\ }\bibfield  {title}
  {\bibinfo {title} {Polymeric quasicrystal: Mesoscopic quasicrystalline tiling
  in {$ABC$} star polymers},\ }\href
  {https://doi.org/10.1103/PhysRevLett.98.195502} {\bibfield  {journal}
  {\bibinfo  {journal} {Phys. Rev. Lett.}\ }\textbf {\bibinfo {volume} {98}},\
  \bibinfo {pages} {195502} (\bibinfo {year} {2007})}\BibitemShut {NoStop}%
\bibitem [{\citenamefont {Talapin}\ \emph {et~al.}(2009)\citenamefont
  {Talapin}, \citenamefont {Shevchenko}, \citenamefont {Bodnarchuk},
  \citenamefont {Ye}, \citenamefont {Chen},\ and\ \citenamefont
  {Murray}}]{Talapin2009}%
  \BibitemOpen
  \bibfield  {author} {\bibinfo {author} {\bibfnamefont {D.~V.}\ \bibnamefont
  {Talapin}}, \bibinfo {author} {\bibfnamefont {E.~V.}\ \bibnamefont
  {Shevchenko}}, \bibinfo {author} {\bibfnamefont {M.~I.}\ \bibnamefont
  {Bodnarchuk}}, \bibinfo {author} {\bibfnamefont {X.~C.}\ \bibnamefont {Ye}},
  \bibinfo {author} {\bibfnamefont {J.}~\bibnamefont {Chen}},\ and\ \bibinfo
  {author} {\bibfnamefont {C.~B.}\ \bibnamefont {Murray}},\ }\bibfield  {title}
  {\bibinfo {title} {Quasicrystalline order in self-assembled binary
  nanoparticle superlattices},\ }\href {https://doi.org/10.1038/nature08439}
  {\bibfield  {journal} {\bibinfo  {journal} {Nature}\ }\textbf {\bibinfo
  {volume} {461}},\ \bibinfo {pages} {964} (\bibinfo {year}
  {2009})}\BibitemShut {NoStop}%
\bibitem [{\citenamefont {Dotera}(2011)}]{Dotera2011}%
  \BibitemOpen
  \bibfield  {author} {\bibinfo {author} {\bibfnamefont {T.}~\bibnamefont
  {Dotera}},\ }\bibfield  {title} {\bibinfo {title} {Quasicrystals in soft
  matter},\ }\href {https://doi.org/10.1002/ijch.201100146} {\bibfield
  {journal} {\bibinfo  {journal} {Isr. J. Chem.}\ }\textbf {\bibinfo {volume}
  {51}},\ \bibinfo {pages} {1197} (\bibinfo {year} {2011})}\BibitemShut
  {NoStop}%
\bibitem [{\citenamefont {Fischer}\ \emph {et~al.}(2011)\citenamefont
  {Fischer}, \citenamefont {Exner}, \citenamefont {Zielske}, \citenamefont
  {Perlich}, \citenamefont {Deloudi}, \citenamefont {Steurer}, \citenamefont
  {Lindner},\ and\ \citenamefont {Forster}}]{Fischer2011}%
  \BibitemOpen
  \bibfield  {author} {\bibinfo {author} {\bibfnamefont {S.}~\bibnamefont
  {Fischer}}, \bibinfo {author} {\bibfnamefont {A.}~\bibnamefont {Exner}},
  \bibinfo {author} {\bibfnamefont {K.}~\bibnamefont {Zielske}}, \bibinfo
  {author} {\bibfnamefont {J.}~\bibnamefont {Perlich}}, \bibinfo {author}
  {\bibfnamefont {S.}~\bibnamefont {Deloudi}}, \bibinfo {author} {\bibfnamefont
  {W.}~\bibnamefont {Steurer}}, \bibinfo {author} {\bibfnamefont
  {P.}~\bibnamefont {Lindner}},\ and\ \bibinfo {author} {\bibfnamefont
  {S.}~\bibnamefont {Forster}},\ }\bibfield  {title} {\bibinfo {title}
  {Colloidal quasicrystals with 12-fold and 18-fold diffraction symmetry},\
  }\href {https://doi.org/10.1073/pnas.1008695108} {\bibfield  {journal}
  {\bibinfo  {journal} {Proc. Natl. Acad. Sci. U.S.A.}\ }\textbf {\bibinfo
  {volume} {108}},\ \bibinfo {pages} {1810} (\bibinfo {year}
  {2011})}\BibitemShut {NoStop}%
\bibitem [{\citenamefont {Xiao}\ \emph {et~al.}(2012)\citenamefont {Xiao},
  \citenamefont {Fujita}, \citenamefont {Miyasaka}, \citenamefont {Sakamoto},\
  and\ \citenamefont {Terasaki}}]{Xiao2012}%
  \BibitemOpen
  \bibfield  {author} {\bibinfo {author} {\bibfnamefont {C.~H.}\ \bibnamefont
  {Xiao}}, \bibinfo {author} {\bibfnamefont {N.}~\bibnamefont {Fujita}},
  \bibinfo {author} {\bibfnamefont {K.}~\bibnamefont {Miyasaka}}, \bibinfo
  {author} {\bibfnamefont {Y.}~\bibnamefont {Sakamoto}},\ and\ \bibinfo
  {author} {\bibfnamefont {O.}~\bibnamefont {Terasaki}},\ }\bibfield  {title}
  {\bibinfo {title} {Dodecagonal tiling in mesoporous silica},\ }\href
  {https://doi.org/10.1038/nature11230} {\bibfield  {journal} {\bibinfo
  {journal} {Nature}\ }\textbf {\bibinfo {volume} {487}},\ \bibinfo {pages}
  {349} (\bibinfo {year} {2012})}\BibitemShut {NoStop}%
\bibitem [{\citenamefont {Ishimasa}(2011)}]{Ishimasa2011}%
  \BibitemOpen
  \bibfield  {author} {\bibinfo {author} {\bibfnamefont {T.}~\bibnamefont
  {Ishimasa}},\ }\bibfield  {title} {\bibinfo {title} {Dodecagonal
  quasicrystals still in progress},\ }\href
  {https://doi.org/10.1002/ijch.201100134} {\bibfield  {journal} {\bibinfo
  {journal} {Isr. J. Chem.}\ }\textbf {\bibinfo {volume} {51}},\ \bibinfo
  {pages} {1216} (\bibinfo {year} {2011})}\BibitemShut {NoStop}%
\bibitem [{\citenamefont {F{\"o}rster}\ \emph {et~al.}(2016)\citenamefont
  {F{\"o}rster}, \citenamefont {Trautmann}, \citenamefont {Roy}, \citenamefont
  {Adeagbo}, \citenamefont {Zollner}, \citenamefont {Hammer}, \citenamefont
  {Schumann}, \citenamefont {Meinel}, \citenamefont {Nayak}, \citenamefont
  {Mohseni}, \citenamefont {Hergert}, \citenamefont {Meyerheim},\ and\
  \citenamefont {Widdra}}]{Forster2016}%
  \BibitemOpen
  \bibfield  {author} {\bibinfo {author} {\bibfnamefont {S.}~\bibnamefont
  {F{\"o}rster}}, \bibinfo {author} {\bibfnamefont {M.}~\bibnamefont
  {Trautmann}}, \bibinfo {author} {\bibfnamefont {S.}~\bibnamefont {Roy}},
  \bibinfo {author} {\bibfnamefont {W.~A.}\ \bibnamefont {Adeagbo}}, \bibinfo
  {author} {\bibfnamefont {E.~M.}\ \bibnamefont {Zollner}}, \bibinfo {author}
  {\bibfnamefont {R.}~\bibnamefont {Hammer}}, \bibinfo {author} {\bibfnamefont
  {F.~O.}\ \bibnamefont {Schumann}}, \bibinfo {author} {\bibfnamefont
  {K.}~\bibnamefont {Meinel}}, \bibinfo {author} {\bibfnamefont {S.~K.}\
  \bibnamefont {Nayak}}, \bibinfo {author} {\bibfnamefont {K.}~\bibnamefont
  {Mohseni}}, \bibinfo {author} {\bibfnamefont {W.}~\bibnamefont {Hergert}},
  \bibinfo {author} {\bibfnamefont {H.~L.}\ \bibnamefont {Meyerheim}},\ and\
  \bibinfo {author} {\bibfnamefont {W.}~\bibnamefont {Widdra}},\ }\bibfield
  {title} {\bibinfo {title} {Observation and structure determination of an
  oxide quasicrystal approximant},\ }\href
  {https://doi.org/10.1103/PhysRevLett.117.095501} {\bibfield  {journal}
  {\bibinfo  {journal} {Phys. Rev. Lett.}\ }\textbf {\bibinfo {volume} {117}},\
  \bibinfo {pages} {095501} (\bibinfo {year} {2016})}\BibitemShut {NoStop}%
\bibitem [{\citenamefont {Ye}\ \emph {et~al.}(2017)\citenamefont {Ye},
  \citenamefont {Chen}, \citenamefont {Irrgang}, \citenamefont {Engel},
  \citenamefont {Dong}, \citenamefont {Glotzer},\ and\ \citenamefont
  {Murray}}]{Ye2017}%
  \BibitemOpen
  \bibfield  {author} {\bibinfo {author} {\bibfnamefont {X.}~\bibnamefont
  {Ye}}, \bibinfo {author} {\bibfnamefont {J.}~\bibnamefont {Chen}}, \bibinfo
  {author} {\bibfnamefont {M.~E.}\ \bibnamefont {Irrgang}}, \bibinfo {author}
  {\bibfnamefont {M.}~\bibnamefont {Engel}}, \bibinfo {author} {\bibfnamefont
  {A.}~\bibnamefont {Dong}}, \bibinfo {author} {\bibfnamefont {S.~C.}\
  \bibnamefont {Glotzer}},\ and\ \bibinfo {author} {\bibfnamefont {C.~B.}\
  \bibnamefont {Murray}},\ }\bibfield  {title} {\bibinfo {title}
  {Quasicrystalline nanocrystal superlattice with partial matching rules},\
  }\href {https://doi.org/10.1038/nmat4759} {\bibfield  {journal} {\bibinfo
  {journal} {Nature Mat.}\ }\textbf {\bibinfo {volume} {16}},\ \bibinfo {pages}
  {214} (\bibinfo {year} {2017})}\BibitemShut {NoStop}%
\bibitem [{\citenamefont {Jayaraman}\ \emph {et~al.}(2021)\citenamefont
  {Jayaraman}, \citenamefont {Baez-Cotto}, \citenamefont {Mann},\ and\
  \citenamefont {Mahanthappa}}]{Jayaraman2021}%
  \BibitemOpen
  \bibfield  {author} {\bibinfo {author} {\bibfnamefont {A.}~\bibnamefont
  {Jayaraman}}, \bibinfo {author} {\bibfnamefont {C.~M.}\ \bibnamefont
  {Baez-Cotto}}, \bibinfo {author} {\bibfnamefont {T.~J.}\ \bibnamefont
  {Mann}},\ and\ \bibinfo {author} {\bibfnamefont {M.~K.}\ \bibnamefont
  {Mahanthappa}},\ }\bibfield  {title} {\bibinfo {title} {Dodecagonal
  quasicrystals of oil-swollen ionic surfactant micelles},\ }\href
  {https://doi.org/10.1073/pnas.2101598118} {\bibfield  {journal} {\bibinfo
  {journal} {Proc. Natl. Acad. Sci. U.S.A.}\ }\textbf {\bibinfo {volume}
  {118}},\ \bibinfo {pages} {e2101598118} (\bibinfo {year} {2021})}\BibitemShut
  {NoStop}%
\bibitem [{\citenamefont {Dotera}\ \emph {et~al.}(2017)\citenamefont {Dotera},
  \citenamefont {Bekku},\ and\ \citenamefont {Ziherl}}]{Dotera2017}%
  \BibitemOpen
  \bibfield  {author} {\bibinfo {author} {\bibfnamefont {T.}~\bibnamefont
  {Dotera}}, \bibinfo {author} {\bibfnamefont {S.}~\bibnamefont {Bekku}},\ and\
  \bibinfo {author} {\bibfnamefont {P.}~\bibnamefont {Ziherl}},\ }\bibfield
  {title} {\bibinfo {title} {Bronze-mean hexagonal quasicrystal},\ }\href
  {https://doi.org/10.1038/nmat4963} {\bibfield  {journal} {\bibinfo  {journal}
  {Nature Mat.}\ }\textbf {\bibinfo {volume} {16}},\ \bibinfo {pages} {987}
  (\bibinfo {year} {2017})}\BibitemShut {NoStop}%
\bibitem [{\citenamefont {Penrose}(1974)}]{Penrose1974}%
  \BibitemOpen
  \bibfield  {author} {\bibinfo {author} {\bibfnamefont {R.}~\bibnamefont
  {Penrose}},\ }\bibfield  {title} {\bibinfo {title} {The role of aesthetics in
  pure and applied mathematical research},\ }\href@noop {} {\bibfield
  {journal} {\bibinfo  {journal} {Bull. Inst. Math. Appl.}\ }\textbf {\bibinfo
  {volume} {10}},\ \bibinfo {pages} {266} (\bibinfo {year} {1974})}\BibitemShut
  {NoStop}%
\bibitem [{\citenamefont {Lifshitz}\ and\ \citenamefont
  {Diamant}(2007)}]{Lifshitz2007a}%
  \BibitemOpen
  \bibfield  {author} {\bibinfo {author} {\bibfnamefont {R.}~\bibnamefont
  {Lifshitz}}\ and\ \bibinfo {author} {\bibfnamefont {H.}~\bibnamefont
  {Diamant}},\ }\bibfield  {title} {\bibinfo {title} {Soft quasicrystals --
  {W}hy are they stable?},\ }\href {https://doi.org/10.1080/14786430701358673}
  {\bibfield  {journal} {\bibinfo  {journal} {Philos. Mag.}\ }\textbf {\bibinfo
  {volume} {87}},\ \bibinfo {pages} {3021} (\bibinfo {year}
  {2007})}\BibitemShut {NoStop}%
\bibitem [{\citenamefont {Nakakura}\ \emph {et~al.}(2019)\citenamefont
  {Nakakura}, \citenamefont {Ziherl}, \citenamefont {Matsuzawa},\ and\
  \citenamefont {Dotera}}]{Nakakura2019}%
  \BibitemOpen
  \bibfield  {author} {\bibinfo {author} {\bibfnamefont {J.}~\bibnamefont
  {Nakakura}}, \bibinfo {author} {\bibfnamefont {P.}~\bibnamefont {Ziherl}},
  \bibinfo {author} {\bibfnamefont {J.}~\bibnamefont {Matsuzawa}},\ and\
  \bibinfo {author} {\bibfnamefont {T.}~\bibnamefont {Dotera}},\ }\bibfield
  {title} {\bibinfo {title} {Metallic-mean quasicrystals as aperiodic
  approximants of periodic crystals},\ }\href
  {https://doi.org/10.1038/s41467-019-12147-z} {\bibfield  {journal} {\bibinfo
  {journal} {Nature Comm.}\ }\textbf {\bibinfo {volume} {10}},\ \bibinfo
  {pages} {4235} (\bibinfo {year} {2019})}\BibitemShut {NoStop}%
\bibitem [{\citenamefont {Cross}\ and\ \citenamefont
  {Hohenberg}(1993)}]{Cross1993}%
  \BibitemOpen
  \bibfield  {author} {\bibinfo {author} {\bibfnamefont {M.~C.}\ \bibnamefont
  {Cross}}\ and\ \bibinfo {author} {\bibfnamefont {P.~C.}\ \bibnamefont
  {Hohenberg}},\ }\bibfield  {title} {\bibinfo {title} {Pattern formation
  outside of equilibrium},\ }\href {https://doi.org/10.1103/RevModPhys.65.851}
  {\bibfield  {journal} {\bibinfo  {journal} {Rev. Mod. Phys.}\ }\textbf
  {\bibinfo {volume} {65}},\ \bibinfo {pages} {851} (\bibinfo {year}
  {1993})}\BibitemShut {NoStop}%
\bibitem [{\citenamefont {Hoyle}(2006)}]{Hoyle2006}%
  \BibitemOpen
  \bibfield  {author} {\bibinfo {author} {\bibfnamefont {R.~B.}\ \bibnamefont
  {Hoyle}},\ }\href {https://doi.org/10.1017/CBO9780511616051} {\emph {\bibinfo
  {title} {Pattern Formation: an Introduction to Methods}}}\ (\bibinfo
  {publisher} {Cambridge University Press},\ \bibinfo {address} {Cambridge},\
  \bibinfo {year} {2006})\BibitemShut {NoStop}%
\bibitem [{\citenamefont {Marconi}\ and\ \citenamefont
  {Tarazona}(1999)}]{Marconi1999}%
  \BibitemOpen
  \bibfield  {author} {\bibinfo {author} {\bibfnamefont {U.~M.~B.}\
  \bibnamefont {Marconi}}\ and\ \bibinfo {author} {\bibfnamefont
  {P.}~\bibnamefont {Tarazona}},\ }\bibfield  {title} {\bibinfo {title}
  {Dynamic density functional theory of fluids},\ }\href
  {https://doi.org/10.1063/1.478705} {\bibfield  {journal} {\bibinfo  {journal}
  {J. Chem. Phys.}\ }\textbf {\bibinfo {volume} {110}},\ \bibinfo {pages}
  {8032} (\bibinfo {year} {1999})}\BibitemShut {NoStop}%
\bibitem [{\citenamefont {Archer}\ and\ \citenamefont
  {Evans}(2004)}]{Archer2004a}%
  \BibitemOpen
  \bibfield  {author} {\bibinfo {author} {\bibfnamefont {A.~J.}\ \bibnamefont
  {Archer}}\ and\ \bibinfo {author} {\bibfnamefont {R.}~\bibnamefont {Evans}},\
  }\bibfield  {title} {\bibinfo {title} {Dynamical density functional theory
  and its application to spinodal decomposition},\ }\href
  {https://doi.org/10.1063/1.1778374} {\bibfield  {journal} {\bibinfo
  {journal} {J. Chem. Phys.}\ }\textbf {\bibinfo {volume} {121}},\ \bibinfo
  {pages} {4246} (\bibinfo {year} {2004})}\BibitemShut {NoStop}%
\bibitem [{\citenamefont {Hansen}\ and\ \citenamefont
  {McDonald}(2013)}]{Hansen2013}%
  \BibitemOpen
  \bibfield  {author} {\bibinfo {author} {\bibfnamefont {J.-P.}\ \bibnamefont
  {Hansen}}\ and\ \bibinfo {author} {\bibfnamefont {I.~R.}\ \bibnamefont
  {McDonald}},\ }\href {https://doi.org/10.1016/C2010-0-66723-X} {\emph
  {\bibinfo {title} {Theory of Simple Liquids with Applications to Soft Matter:
  Fourth Edition}}}\ (\bibinfo  {publisher} {Elsevier},\ \bibinfo {address}
  {Oxford},\ \bibinfo {year} {2013})\BibitemShut {NoStop}%
\bibitem [{\citenamefont {te~Vrugt}\ \emph {et~al.}(2020)\citenamefont
  {te~Vrugt}, \citenamefont {L{\"o}wen},\ and\ \citenamefont
  {Wittkowski}}]{teVrugt2020}%
  \BibitemOpen
  \bibfield  {author} {\bibinfo {author} {\bibfnamefont {M.}~\bibnamefont
  {te~Vrugt}}, \bibinfo {author} {\bibfnamefont {H.}~\bibnamefont
  {L{\"o}wen}},\ and\ \bibinfo {author} {\bibfnamefont {R.}~\bibnamefont
  {Wittkowski}},\ }\bibfield  {title} {\bibinfo {title} {Classical dynamical
  density functional theory: from fundamentals to applications},\ }\href
  {https://doi.org/10.1080/00018732.2020.1854965} {\bibfield  {journal}
  {\bibinfo  {journal} {Adv. Phys.}\ }\textbf {\bibinfo {volume} {69}},\
  \bibinfo {pages} {121} (\bibinfo {year} {2020})}\BibitemShut {NoStop}%
\bibitem [{\citenamefont {Alexander}\ and\ \citenamefont
  {McTague}(1978)}]{Alexander1978}%
  \BibitemOpen
  \bibfield  {author} {\bibinfo {author} {\bibfnamefont {S.}~\bibnamefont
  {Alexander}}\ and\ \bibinfo {author} {\bibfnamefont {J.}~\bibnamefont
  {McTague}},\ }\bibfield  {title} {\bibinfo {title} {Should all crystals be
  bcc? {L}andau theory of solidification and crystal nucleation},\ }\href
  {https://doi.org/10.1103/PhysRevLett.41.702} {\bibfield  {journal} {\bibinfo
  {journal} {Phys. Rev. Lett.}\ }\textbf {\bibinfo {volume} {41}},\ \bibinfo
  {pages} {702} (\bibinfo {year} {1978})}\BibitemShut {NoStop}%
\bibitem [{\citenamefont {Bak}(1985)}]{Bak1985}%
  \BibitemOpen
  \bibfield  {author} {\bibinfo {author} {\bibfnamefont {P.}~\bibnamefont
  {Bak}},\ }\bibfield  {title} {\bibinfo {title} {Phenomenological theory of
  icosahedral incommensurate (``quasiperiodic'') order in {M}n-{A}l alloys},\
  }\href {https://doi.org/10.1103/PhysRevLett.54.1517} {\bibfield  {journal}
  {\bibinfo  {journal} {Phys. Rev. Lett.}\ }\textbf {\bibinfo {volume} {54}},\
  \bibinfo {pages} {1517} (\bibinfo {year} {1985})}\BibitemShut {NoStop}%
\bibitem [{\citenamefont {Lifshitz}\ and\ \citenamefont
  {Petrich}(1997)}]{Lifshitz1997}%
  \BibitemOpen
  \bibfield  {author} {\bibinfo {author} {\bibfnamefont {R.}~\bibnamefont
  {Lifshitz}}\ and\ \bibinfo {author} {\bibfnamefont {D.~M.}\ \bibnamefont
  {Petrich}},\ }\bibfield  {title} {\bibinfo {title} {Theoretical model for
  {F}araday waves with multiple-frequency forcing},\ }\href
  {https://doi.org/10.1103/PhysRevLett.79.1261} {\bibfield  {journal} {\bibinfo
   {journal} {Phys. Rev. Lett.}\ }\textbf {\bibinfo {volume} {79}},\ \bibinfo
  {pages} {1261} (\bibinfo {year} {1997})}\BibitemShut {NoStop}%
\bibitem [{\citenamefont {Rucklidge}\ \emph {et~al.}(2012)\citenamefont
  {Rucklidge}, \citenamefont {Silber},\ and\ \citenamefont
  {Skeldon}}]{Rucklidge2012}%
  \BibitemOpen
  \bibfield  {author} {\bibinfo {author} {\bibfnamefont {A.~M.}\ \bibnamefont
  {Rucklidge}}, \bibinfo {author} {\bibfnamefont {M.}~\bibnamefont {Silber}},\
  and\ \bibinfo {author} {\bibfnamefont {A.~C.}\ \bibnamefont {Skeldon}},\
  }\bibfield  {title} {\bibinfo {title} {Three-wave interactions and
  spatiotemporal chaos},\ }\href
  {https://doi.org/10.1103/PhysRevLett.108.074504} {\bibfield  {journal}
  {\bibinfo  {journal} {Phys. Rev. Lett.}\ }\textbf {\bibinfo {volume} {108}},\
  \bibinfo {pages} {074504} (\bibinfo {year} {2012})}\BibitemShut {NoStop}%
\bibitem [{\citenamefont {Archer}\ \emph {et~al.}(2013)\citenamefont {Archer},
  \citenamefont {Rucklidge},\ and\ \citenamefont {Knobloch}}]{Archer2013}%
  \BibitemOpen
  \bibfield  {author} {\bibinfo {author} {\bibfnamefont {A.~J.}\ \bibnamefont
  {Archer}}, \bibinfo {author} {\bibfnamefont {A.~M.}\ \bibnamefont
  {Rucklidge}},\ and\ \bibinfo {author} {\bibfnamefont {E.}~\bibnamefont
  {Knobloch}},\ }\bibfield  {title} {\bibinfo {title} {Quasicrystalline order
  and a crystal-liquid state in a soft-core fluid},\ }\href
  {https://doi.org/10.1103/PhysRevLett.111.165501} {\bibfield  {journal}
  {\bibinfo  {journal} {Phys. Rev. Lett.}\ }\textbf {\bibinfo {volume} {111}},\
  \bibinfo {pages} {165501} (\bibinfo {year} {2013})}\BibitemShut {NoStop}%
\bibitem [{\citenamefont {Ratliff}\ \emph {et~al.}(2019)\citenamefont
  {Ratliff}, \citenamefont {Archer}, \citenamefont {Subramanian},\ and\
  \citenamefont {Rucklidge}}]{Ratliff2019}%
  \BibitemOpen
  \bibfield  {author} {\bibinfo {author} {\bibfnamefont {D.~J.}\ \bibnamefont
  {Ratliff}}, \bibinfo {author} {\bibfnamefont {A.~J.}\ \bibnamefont {Archer}},
  \bibinfo {author} {\bibfnamefont {P.}~\bibnamefont {Subramanian}},\ and\
  \bibinfo {author} {\bibfnamefont {A.~M.}\ \bibnamefont {Rucklidge}},\
  }\bibfield  {title} {\bibinfo {title} {Which wave numbers determine the
  thermodynamic stability of soft matter quasicrystals?},\ }\href
  {https://doi.org/10.1103/PhysRevLett.123.148004} {\bibfield  {journal}
  {\bibinfo  {journal} {Phys. Rev. Lett.}\ }\textbf {\bibinfo {volume} {123}},\
  \bibinfo {pages} {148004} (\bibinfo {year} {2019})}\BibitemShut {NoStop}%
\bibitem [{\citenamefont {Castelino}\ \emph {et~al.}(2020)\citenamefont
  {Castelino}, \citenamefont {Ratliff}, \citenamefont {Rucklidge},
  \citenamefont {Subramanian},\ and\ \citenamefont {Topaz}}]{Castelino2020}%
  \BibitemOpen
  \bibfield  {author} {\bibinfo {author} {\bibfnamefont {J.~K.}\ \bibnamefont
  {Castelino}}, \bibinfo {author} {\bibfnamefont {D.~J.}\ \bibnamefont
  {Ratliff}}, \bibinfo {author} {\bibfnamefont {A.~M.}\ \bibnamefont
  {Rucklidge}}, \bibinfo {author} {\bibfnamefont {P.}~\bibnamefont
  {Subramanian}},\ and\ \bibinfo {author} {\bibfnamefont {C.~M.}\ \bibnamefont
  {Topaz}},\ }\bibfield  {title} {\bibinfo {title} {Spatiotemporal chaos and
  quasipatterns in coupled reaction--diffusion systems},\ }\href
  {https://doi.org/10.1016/j.physd.2020.132475} {\bibfield  {journal} {\bibinfo
   {journal} {Physica D}\ }\textbf {\bibinfo {volume} {407}},\ \bibinfo {pages}
  {132475} (\bibinfo {year} {2020})}\BibitemShut {NoStop}%
\bibitem [{\citenamefont {Evans}(1979)}]{Evans1979a}%
  \BibitemOpen
  \bibfield  {author} {\bibinfo {author} {\bibfnamefont {R.}~\bibnamefont
  {Evans}},\ }\bibfield  {title} {\bibinfo {title} {The nature of the
  liquid-vapour interface and other topics in the statistical mechanics of
  non-uniform, classical fluids},\ }\href
  {https://doi.org/10.1080/00018737900101365} {\bibfield  {journal} {\bibinfo
  {journal} {Adv. Phys.}\ }\textbf {\bibinfo {volume} {28}},\ \bibinfo {pages}
  {143} (\bibinfo {year} {1979})}\BibitemShut {NoStop}%
\bibitem [{\citenamefont {Likos}(2001)}]{Likos2001}%
  \BibitemOpen
  \bibfield  {author} {\bibinfo {author} {\bibfnamefont {C.~N.}\ \bibnamefont
  {Likos}},\ }\bibfield  {title} {\bibinfo {title} {Effective interactions in
  soft condensed matter physics},\ }\href
  {https://doi.org/10.1016/S0370-1573(00)00141-1} {\bibfield  {journal}
  {\bibinfo  {journal} {Phys. Rep.}\ }\textbf {\bibinfo {volume} {348}},\
  \bibinfo {pages} {267} (\bibinfo {year} {2001})}\BibitemShut {NoStop}%
\bibitem [{\citenamefont {Archer}\ \emph {et~al.}(2014)\citenamefont {Archer},
  \citenamefont {Walters}, \citenamefont {Thiele},\ and\ \citenamefont
  {Knobloch}}]{Archer2014}%
  \BibitemOpen
  \bibfield  {author} {\bibinfo {author} {\bibfnamefont {A.~J.}\ \bibnamefont
  {Archer}}, \bibinfo {author} {\bibfnamefont {M.~C.}\ \bibnamefont {Walters}},
  \bibinfo {author} {\bibfnamefont {U.}~\bibnamefont {Thiele}},\ and\ \bibinfo
  {author} {\bibfnamefont {E.}~\bibnamefont {Knobloch}},\ }\bibfield  {title}
  {\bibinfo {title} {Solidification in soft-core fluids: Disordered solids from
  fast solidification fronts},\ }\href
  {https://doi.org/10.1103/PhysRevE.90.042404} {\bibfield  {journal} {\bibinfo
  {journal} {Phys. Rev. E}\ }\textbf {\bibinfo {volume} {90}},\ \bibinfo
  {pages} {042404} (\bibinfo {year} {2014})}\BibitemShut {NoStop}%
\bibitem [{\citenamefont {Walters}\ \emph {et~al.}(2018)\citenamefont
  {Walters}, \citenamefont {Subramanian}, \citenamefont {Archer},\ and\
  \citenamefont {Evans}}]{Walters2018}%
  \BibitemOpen
  \bibfield  {author} {\bibinfo {author} {\bibfnamefont {M.~C.}\ \bibnamefont
  {Walters}}, \bibinfo {author} {\bibfnamefont {P.}~\bibnamefont
  {Subramanian}}, \bibinfo {author} {\bibfnamefont {A.~J.}\ \bibnamefont
  {Archer}},\ and\ \bibinfo {author} {\bibfnamefont {R.}~\bibnamefont
  {Evans}},\ }\bibfield  {title} {\bibinfo {title} {Structural crossover in a
  model fluid exhibiting two length scales: repercussions for quasicrystal
  formation},\ }\href {https://doi.org/10.1103/PhysRevE.98.012606} {\bibfield
  {journal} {\bibinfo  {journal} {Phys. Rev. E}\ }\textbf {\bibinfo {volume}
  {98}},\ \bibinfo {pages} {012606} (\bibinfo {year} {2018})}\BibitemShut
  {NoStop}%
\bibitem [{\citenamefont {Archer}\ \emph {et~al.}(2012)\citenamefont {Archer},
  \citenamefont {Robbins}, \citenamefont {Thiele},\ and\ \citenamefont
  {Knobloch}}]{Archer2012}%
  \BibitemOpen
  \bibfield  {author} {\bibinfo {author} {\bibfnamefont {A.~J.}\ \bibnamefont
  {Archer}}, \bibinfo {author} {\bibfnamefont {M.~J.}\ \bibnamefont {Robbins}},
  \bibinfo {author} {\bibfnamefont {U.}~\bibnamefont {Thiele}},\ and\ \bibinfo
  {author} {\bibfnamefont {E.}~\bibnamefont {Knobloch}},\ }\bibfield  {title}
  {\bibinfo {title} {Solidification fronts in supercooled liquids: {H}ow rapid
  fronts can lead to disordered glassy solids},\ }\href
  {https://doi.org/{10.1103/PhysRevE.86.031603}} {\bibfield  {journal}
  {\bibinfo  {journal} {Phys. Rev. E}\ }\textbf {\bibinfo {volume} {{86}}},\
  \bibinfo {pages} {031603} (\bibinfo {year} {{2012}})}\BibitemShut {NoStop}%
\bibitem [{\citenamefont {Barkan}\ \emph {et~al.}(2014)\citenamefont {Barkan},
  \citenamefont {Engel},\ and\ \citenamefont {Lifshitz}}]{Barkan2014}%
  \BibitemOpen
  \bibfield  {author} {\bibinfo {author} {\bibfnamefont {K.}~\bibnamefont
  {Barkan}}, \bibinfo {author} {\bibfnamefont {M.}~\bibnamefont {Engel}},\ and\
  \bibinfo {author} {\bibfnamefont {R.}~\bibnamefont {Lifshitz}},\ }\bibfield
  {title} {\bibinfo {title} {Controlled self-assembly of periodic and aperiodic
  cluster crystals},\ }\href {https://doi.org/10.1103/PhysRevLett.113.098304}
  {\bibfield  {journal} {\bibinfo  {journal} {Phys. Rev. Lett.}\ }\textbf
  {\bibinfo {volume} {113}},\ \bibinfo {pages} {098304} (\bibinfo {year}
  {2014})}\BibitemShut {NoStop}%
\bibitem [{\citenamefont {Barkan}\ \emph {et~al.}(2011)\citenamefont {Barkan},
  \citenamefont {Diamant},\ and\ \citenamefont {Lifshitz}}]{Barkan2011}%
  \BibitemOpen
  \bibfield  {author} {\bibinfo {author} {\bibfnamefont {K.}~\bibnamefont
  {Barkan}}, \bibinfo {author} {\bibfnamefont {H.}~\bibnamefont {Diamant}},\
  and\ \bibinfo {author} {\bibfnamefont {R.}~\bibnamefont {Lifshitz}},\
  }\bibfield  {title} {\bibinfo {title} {Stability of quasicrystals composed of
  soft isotropic particles},\ }\href
  {https://doi.org/10.1103/PhysRevB.83.172201} {\bibfield  {journal} {\bibinfo
  {journal} {Phys. Rev. B}\ }\textbf {\bibinfo {volume} {83}},\ \bibinfo
  {pages} {172201} (\bibinfo {year} {2011})}\BibitemShut {NoStop}%
\bibitem [{\citenamefont {Zu}\ \emph {et~al.}(2017)\citenamefont {Zu},
  \citenamefont {Tan},\ and\ \citenamefont {Xu}}]{Zu2017}%
  \BibitemOpen
  \bibfield  {author} {\bibinfo {author} {\bibfnamefont {M.}~\bibnamefont
  {Zu}}, \bibinfo {author} {\bibfnamefont {P.}~\bibnamefont {Tan}},\ and\
  \bibinfo {author} {\bibfnamefont {N.}~\bibnamefont {Xu}},\ }\bibfield
  {title} {\bibinfo {title} {Forming quasicrystals by monodisperse soft core
  particles},\ }\href {https://doi.org/10.1038/s41467-017-02316-3} {\bibfield
  {journal} {\bibinfo  {journal} {Nat. Commun.}\ }\textbf {\bibinfo {volume}
  {8}},\ \bibinfo {pages} {2089} (\bibinfo {year} {2017})}\BibitemShut
  {NoStop}%
\bibitem [{\citenamefont {Scacchi}\ \emph {et~al.}(2020)\citenamefont
  {Scacchi}, \citenamefont {Somerville}, \citenamefont {Buzza},\ and\
  \citenamefont {Archer}}]{Scacchi2020}%
  \BibitemOpen
  \bibfield  {author} {\bibinfo {author} {\bibfnamefont {A.}~\bibnamefont
  {Scacchi}}, \bibinfo {author} {\bibfnamefont {W.~R.~C.}\ \bibnamefont
  {Somerville}}, \bibinfo {author} {\bibfnamefont {D.~M.~A.}\ \bibnamefont
  {Buzza}},\ and\ \bibinfo {author} {\bibfnamefont {A.~J.}\ \bibnamefont
  {Archer}},\ }\bibfield  {title} {\bibinfo {title} {Quasicrystal formation in
  binary soft matter mixtures},\ }\href
  {https://doi.org/10.1103/PhysRevResearch.2.032043} {\bibfield  {journal}
  {\bibinfo  {journal} {Phys. Rev. Res.}\ }\textbf {\bibinfo {volume} {2}},\
  \bibinfo {pages} {032043(R)} (\bibinfo {year} {2020})}\BibitemShut {NoStop}%
\bibitem [{\citenamefont {Malescio}\ and\ \citenamefont
  {Sciortino}(2021)}]{Malescio2021}%
  \BibitemOpen
  \bibfield  {author} {\bibinfo {author} {\bibfnamefont {G.}~\bibnamefont
  {Malescio}}\ and\ \bibinfo {author} {\bibfnamefont {F.}~\bibnamefont
  {Sciortino}},\ }\bibfield  {title} {\bibinfo {title} {Self-assembly of
  quasicrystals and their approximants in fluids with bounded repulsive core
  and competing interactions},\ }\bibfield  {journal} {\bibinfo  {journal} {J.
  Mol. Liq.}\ }\href {https://doi.org/10.1016/j.molliq.2021.118209}
  {10.1016/j.molliq.2021.118209} (\bibinfo {year} {2021})\BibitemShut {NoStop}%
\bibitem [{\citenamefont {Barkan}(2015)}]{Barkan2015}%
  \BibitemOpen
  \bibfield  {author} {\bibinfo {author} {\bibfnamefont {K.}~\bibnamefont
  {Barkan}},\ }\emph {\bibinfo {title} {Theory and Simulation of the Self
  Assembly of Soft Quasicrystals}},\ \href@noop {} {Ph.D. thesis},\ \bibinfo
  {school} {Tel Aviv University} (\bibinfo {year} {2015})\BibitemShut {NoStop}%
\bibitem [{\citenamefont {Roth}(2010)}]{Roth2010}%
  \BibitemOpen
  \bibfield  {author} {\bibinfo {author} {\bibfnamefont {R.}~\bibnamefont
  {Roth}},\ }\bibfield  {title} {\bibinfo {title} {Fundamental measure theory
  of hard-sphere mixtures: a review},\ }\href
  {https://doi.org/10.1088/0953-8984/22/6/063102} {\bibfield  {journal}
  {\bibinfo  {journal} {J. Phys.: Condens. Matter}\ }\textbf {\bibinfo {volume}
  {22}},\ \bibinfo {pages} {063102} (\bibinfo {year} {2010})}\BibitemShut
  {NoStop}%
\bibitem [{\citenamefont {Archer}\ \emph {et~al.}(2022)\citenamefont {Archer},
  \citenamefont {Dotera},\ and\ \citenamefont {Rucklidge}}]{ADR2022}%
  \BibitemOpen
  \bibfield  {author} {\bibinfo {author} {\bibfnamefont {A.~J.}~\bibnamefont
  {Archer}}, \bibinfo {author} {\bibfnamefont {T.}~\bibnamefont {Dotera}},\ and\
  \bibinfo {author} {\bibfnamefont {A.~M.}\ \bibnamefont {Rucklidge}},\ }\bibfield
  {title} {\bibinfo {title} {Dataset for ``Rectangle--triangle soft-matter quasicrystals 
  with hexagonal symmetry''},\ }\href
  {https://doi.org/10.5518/1188} {\bibfield  {journal} {\bibinfo
  {journal} {University of Leeds. [Dataset]. https://doi.org/10.5518/1188}\ }\
  (\bibinfo {year} {2022})}\BibitemShut {NoStop}%
\bibitem [{\citenamefont {Rucklidge}\ and\ \citenamefont
  {Silber}(2009)}]{Rucklidge2009}%
  \BibitemOpen
  \bibfield  {author} {\bibinfo {author} {\bibfnamefont {A.~M.}\ \bibnamefont
  {Rucklidge}}\ and\ \bibinfo {author} {\bibfnamefont {M.}~\bibnamefont
  {Silber}},\ }\bibfield  {title} {\bibinfo {title} {Design of parametrically
  forced patterns and quasipatterns},\ }\href
  {https://doi.org/10.1137/080719066} {\bibfield  {journal} {\bibinfo
  {journal} {SIAM J. Appl. Dynam. Syst.}\ }\textbf {\bibinfo {volume} {8}},\
  \bibinfo {pages} {298} (\bibinfo {year} {2009})}\BibitemShut {NoStop}%
\bibitem [{\citenamefont {Iooss}\ and\ \citenamefont
  {Rucklidge}(2022)}]{Iooss2022}%
  \BibitemOpen
  \bibfield  {author} {\bibinfo {author} {\bibfnamefont {G.}~\bibnamefont
  {Iooss}}\ and\ \bibinfo {author} {\bibfnamefont {A.~M.}\ \bibnamefont
  {Rucklidge}},\ }\bibfield  {title} {\bibinfo {title} {Patterns and
  quasipatterns from the superposition of two hexagonal lattices},\ }\href
  {https://doi.org/10.1137/20M1372780} {\bibfield  {journal} {\bibinfo
  {journal} {SIAM J. Appl. Dynam. Syst.}\ }\textbf {\bibinfo {volume} {21}},\
  \bibinfo {pages} {1119} (\bibinfo {year} {2022})}\BibitemShut {NoStop}%
\bibitem [{\citenamefont {Archer}\ \emph {et~al.}(2019)\citenamefont {Archer},
  \citenamefont {Ratliff}, \citenamefont {Rucklidge},\ and\ \citenamefont
  {Subramanian}}]{Archer2019}%
  \BibitemOpen
  \bibfield  {author} {\bibinfo {author} {\bibfnamefont {A.~J.}\ \bibnamefont
  {Archer}}, \bibinfo {author} {\bibfnamefont {D.~J.}\ \bibnamefont {Ratliff}},
  \bibinfo {author} {\bibfnamefont {A.~M.}\ \bibnamefont {Rucklidge}},\ and\
  \bibinfo {author} {\bibfnamefont {P.}~\bibnamefont {Subramanian}},\
  }\bibfield  {title} {\bibinfo {title} {Deriving phase field crystal theory
  from dynamical density functional theory: consequences of the
  approximations},\ }\href {https://doi.org/10.1103/PhysRevE.100.022140}
  {\bibfield  {journal} {\bibinfo  {journal} {Phys. Rev. E}\ }\textbf {\bibinfo
  {volume} {100}},\ \bibinfo {pages} {022140} (\bibinfo {year}
  {2019})}\BibitemShut {NoStop}%
\bibitem [{\citenamefont {Subramanian}\ \emph {et~al.}(2021)\citenamefont
  {Subramanian}, \citenamefont {Ratliff}, \citenamefont {Rucklidge},\ and\
  \citenamefont {Archer}}]{Subramanian2021a}%
  \BibitemOpen
  \bibfield  {author} {\bibinfo {author} {\bibfnamefont {P.}~\bibnamefont
  {Subramanian}}, \bibinfo {author} {\bibfnamefont {D.~J.}\ \bibnamefont
  {Ratliff}}, \bibinfo {author} {\bibfnamefont {A.~M.}\ \bibnamefont
  {Rucklidge}},\ and\ \bibinfo {author} {\bibfnamefont {A.~J.}\ \bibnamefont
  {Archer}},\ }\bibfield  {title} {\bibinfo {title} {Density distribution in
  soft matter crystals and quasicrystals},\ }\href
  {https://doi.org/10.1103/PhysRevLett.126.218003} {\bibfield  {journal}
  {\bibinfo  {journal} {Phys. Rev. Lett.}\ }\textbf {\bibinfo {volume} {126}},\
  \bibinfo {pages} {218003} (\bibinfo {year} {2021})}\BibitemShut {NoStop}%
\bibitem [{\citenamefont {Jiang}\ \emph {et~al.}(2017)\citenamefont {Jiang},
  \citenamefont {Zhang},\ and\ \citenamefont {Shi}}]{Jiang2017}%
  \BibitemOpen
  \bibfield  {author} {\bibinfo {author} {\bibfnamefont {K.}~\bibnamefont
  {Jiang}}, \bibinfo {author} {\bibfnamefont {P.}~\bibnamefont {Zhang}},\ and\
  \bibinfo {author} {\bibfnamefont {A.-C.}\ \bibnamefont {Shi}},\ }\bibfield
  {title} {\bibinfo {title} {Stability of icosahedral quasicrystals in a simple
  model with two-length scales},\ }\href
  {https://doi.org/10.1088/1361-648X/aa586b} {\bibfield  {journal} {\bibinfo
  {journal} {J. Phys.: Condens. Matter}\ }\textbf {\bibinfo {volume} {29}},\
  \bibinfo {pages} {124003} (\bibinfo {year} {2017})}\BibitemShut {NoStop}%
\bibitem [{\citenamefont {Goldman}\ and\ \citenamefont
  {Kelton}(1993)}]{Goldman1993}%
  \BibitemOpen
  \bibfield  {author} {\bibinfo {author} {\bibfnamefont {A.~I.}\ \bibnamefont
  {Goldman}}\ and\ \bibinfo {author} {\bibfnamefont {R.~F.}\ \bibnamefont
  {Kelton}},\ }\bibfield  {title} {\bibinfo {title} {Quasicrystals and
  crystalline approximants},\ }\href
  {https://doi.org/10.1103/RevModPhys.65.213} {\bibfield  {journal} {\bibinfo
  {journal} {Rev, Mod. Phys.}\ }\textbf {\bibinfo {volume} {65}},\ \bibinfo
  {pages} {213} (\bibinfo {year} {1993})}\BibitemShut {NoStop}%
\bibitem [{\citenamefont {Dionne}\ \emph {et~al.}(1997)\citenamefont {Dionne},
  \citenamefont {Silber},\ and\ \citenamefont {Skeldon}}]{Dionne1997}%
  \BibitemOpen
  \bibfield  {author} {\bibinfo {author} {\bibfnamefont {B.}~\bibnamefont
  {Dionne}}, \bibinfo {author} {\bibfnamefont {M.}~\bibnamefont {Silber}},\
  and\ \bibinfo {author} {\bibfnamefont {A.~C.}\ \bibnamefont {Skeldon}},\
  }\bibfield  {title} {\bibinfo {title} {Stability results for steady,
  spatially periodic planforms},\ }\href
  {https://doi.org/10.1088/0951-7715/10/2/002} {\bibfield  {journal} {\bibinfo
  {journal} {Nonlinearity}\ }\textbf {\bibinfo {volume} {10}},\ \bibinfo
  {pages} {321} (\bibinfo {year} {1997})}\BibitemShut {NoStop}%
\bibitem [{\citenamefont {Levine}\ and\ \citenamefont
  {Steinhardt}(1984)}]{Levine1984}%
  \BibitemOpen
  \bibfield  {author} {\bibinfo {author} {\bibfnamefont {D.}~\bibnamefont
  {Levine}}\ and\ \bibinfo {author} {\bibfnamefont {P.~J.}\ \bibnamefont
  {Steinhardt}},\ }\bibfield  {title} {\bibinfo {title} {Quasicrystals: a new
  class of ordered structures},\ }\href
  {https://doi.org/10.1103/PhysRevLett.53.2477} {\bibfield  {journal} {\bibinfo
   {journal} {Phys. Rev. Lett.}\ }\textbf {\bibinfo {volume} {53}},\ \bibinfo
  {pages} {2477} (\bibinfo {year} {1984})}\BibitemShut {NoStop}%
\bibitem [{\citenamefont {Mladek}\ \emph {et~al.}(2008)\citenamefont {Mladek},
  \citenamefont {Charbonneau}, \citenamefont {Likos}, \citenamefont {Frenkel},\
  and\ \citenamefont {Kahl}}]{Mladek2008}%
  \BibitemOpen
  \bibfield  {author} {\bibinfo {author} {\bibfnamefont {B.~M.}\ \bibnamefont
  {Mladek}}, \bibinfo {author} {\bibfnamefont {P.}~\bibnamefont {Charbonneau}},
  \bibinfo {author} {\bibfnamefont {C.~N.}\ \bibnamefont {Likos}}, \bibinfo
  {author} {\bibfnamefont {D.}~\bibnamefont {Frenkel}},\ and\ \bibinfo {author}
  {\bibfnamefont {G.}~\bibnamefont {Kahl}},\ }\bibfield  {title} {\bibinfo
  {title} {Multiple occupancy crystals formed by purely repulsive soft
  particles},\ }\href {https://doi.org/10.1088/0953-8984/20/49/494245}
  {\bibfield  {journal} {\bibinfo  {journal} {J. Phys.: Condens. Matter}\
  }\textbf {\bibinfo {volume} {20}},\ \bibinfo {pages} {494245} (\bibinfo
  {year} {2008})}\BibitemShut {NoStop}%
\bibitem [{\citenamefont {Lenz}\ \emph {et~al.}(2012)\citenamefont {Lenz},
  \citenamefont {Blaak}, \citenamefont {Likos},\ and\ \citenamefont
  {Mladek}}]{Lenz2012}%
  \BibitemOpen
  \bibfield  {author} {\bibinfo {author} {\bibfnamefont {D.~A.}\ \bibnamefont
  {Lenz}}, \bibinfo {author} {\bibfnamefont {R.}~\bibnamefont {Blaak}},
  \bibinfo {author} {\bibfnamefont {C.~N.}\ \bibnamefont {Likos}},\ and\
  \bibinfo {author} {\bibfnamefont {B.~M.}\ \bibnamefont {Mladek}},\ }\bibfield
   {title} {\bibinfo {title} {Microscopically resolved simulations prove the
  existence of soft cluster crystals},\ }\href
  {https://doi.org/10.1103/PhysRevLett.109.228301} {\bibfield  {journal}
  {\bibinfo  {journal} {Phys. Rev. Lett.}\ }\textbf {\bibinfo {volume} {109}},\
  \bibinfo {pages} {228301} (\bibinfo {year} {2012})}\BibitemShut {NoStop}%
\bibitem [{\citenamefont {Frettl\"oh}\ \emph {et~al.}(2021)\citenamefont
  {Frettl\"oh}, \citenamefont {Harriss},\ and\ \citenamefont
  {G\"ahler}}]{Frettloh2021}%
  \BibitemOpen
  \bibfield  {author} {\bibinfo {author} {\bibfnamefont {D.}~\bibnamefont
  {Frettl\"oh}}, \bibinfo {author} {\bibfnamefont {E.}~\bibnamefont
  {Harriss}},\ and\ \bibinfo {author} {\bibfnamefont {F.}~\bibnamefont
  {G\"ahler}},\ }\href {https://tilings.math.uni-bielefeld.de/} {\bibinfo
  {title} {Tilings {E}ncyclopedia}} (\bibinfo {year} {2021})\BibitemShut
  {NoStop}%
\bibitem [{\citenamefont {Zeller}\ and\ \citenamefont
  {G{\"u}nther}(2014)}]{Zeller2014}%
  \BibitemOpen
  \bibfield  {author} {\bibinfo {author} {\bibfnamefont {P.}~\bibnamefont
  {Zeller}}\ and\ \bibinfo {author} {\bibfnamefont {S.}~\bibnamefont
  {G{\"u}nther}},\ }\bibfield  {title} {\bibinfo {title} {What are the possible
  moir{\'e} patterns of graphene on hexagonally packed surfaces? {U}niversal
  solution for hexagonal coincidence lattices, derived by a geometric
  construction},\ }\href {https://doi.org/10.1088/1367-2630/16/8/083028}
  {\bibfield  {journal} {\bibinfo  {journal} {New J. Phys.}\ }\textbf {\bibinfo
  {volume} {16}},\ \bibinfo {pages} {083028} (\bibinfo {year}
  {2014})}\BibitemShut {NoStop}%
\end{thebibliography}

%

\end{document}